\documentclass[aps,prb,twocolumn,showpacs,floatfix,longbibliography,superscriptaddress]{revtex4-1} 
\usepackage{amsfonts}
\usepackage{graphicx}
\usepackage{amsmath}
\usepackage{times}
\usepackage{amssymb}
\usepackage{color}
\usepackage{subfigure}
\usepackage[ colorlinks, bookmarks=false, citecolor=blue, linkcolor=red, urlcolor=magenta]{hyperref}
\usepackage[english]{babel}
\usepackage[dvipsnames]{xcolor}
\usepackage[normalem]{ulem}

\newcommand{\diagram}[1]{\vcenter{\hbox{\includegraphics[scale=0.035]{./#1.pdf}}}}

\newcommand{\diagrammB}[1]{\vcenter{\hbox{\includegraphics[scale=0.028]{./#1.pdf}}}}
\newcommand{\diagrammZ}[1]{\vcenter{\hbox{\includegraphics[scale=0.029]{./#1.pdf}}}}
\newcommand{\diagrammM}[1]{\vcenter{\hbox{\includegraphics[scale=0.0301]{./#1.pdf}}}}

\newcommand{\Uten}{\includegraphics[scale=0.005,height=1em]{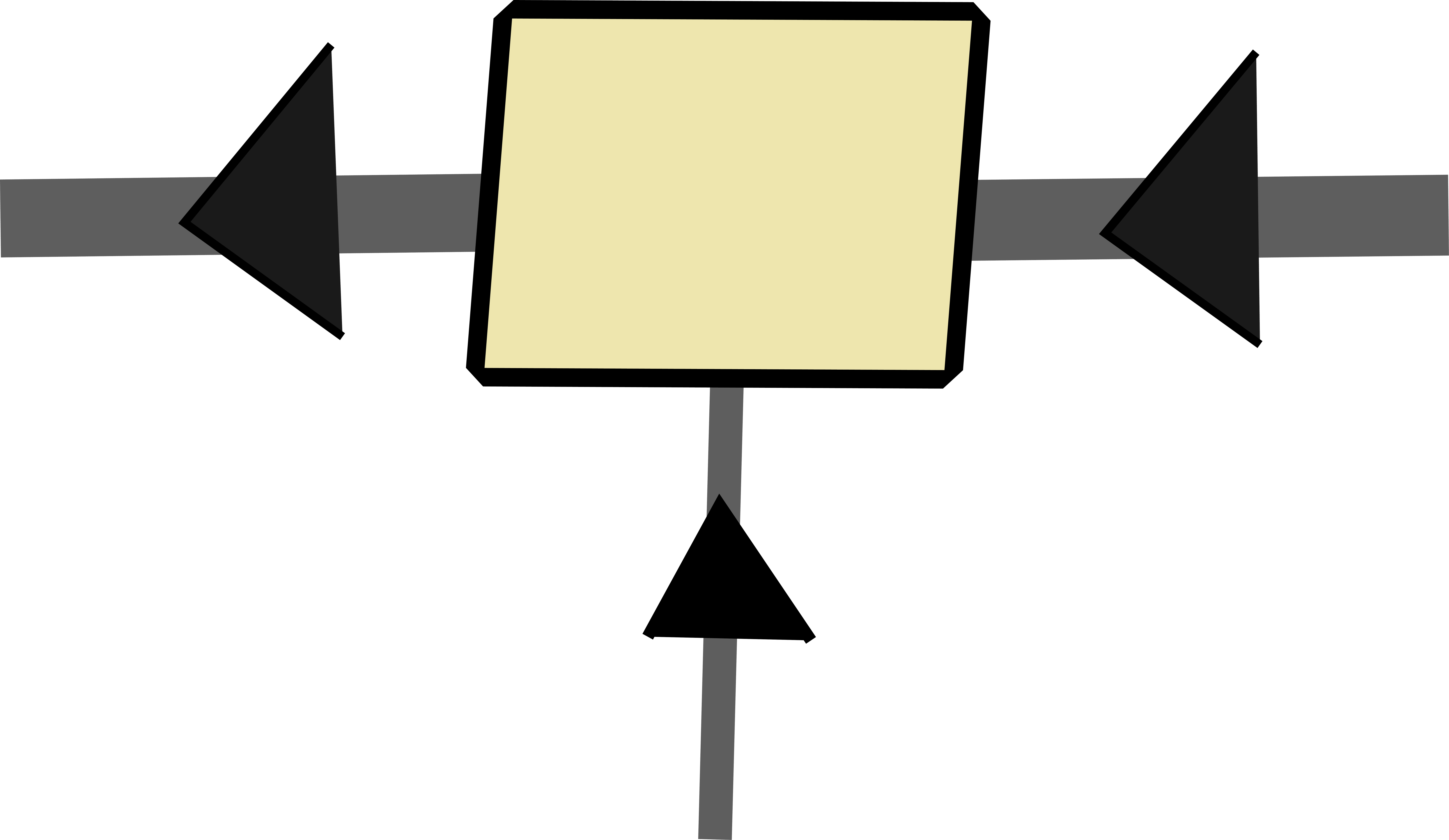}}
\newcommand{\mpsten}{\includegraphics[scale=0.005 ]{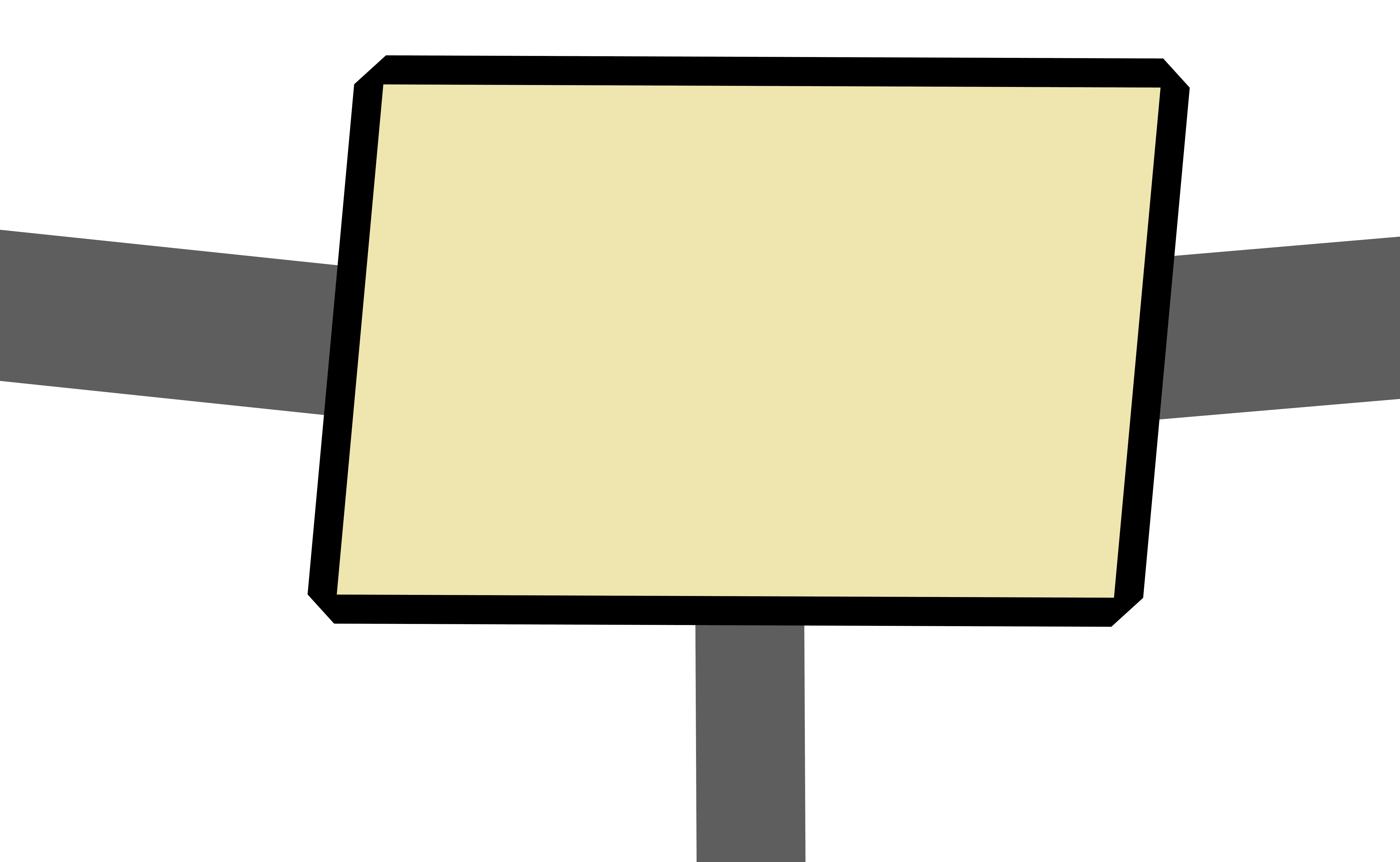}}
\newcommand{\Uiden}{\includegraphics[scale=0.0043]{U_iden.pdf}}
\newcommand{\Uidenr}{\includegraphics[scale=0.0043]{U_iden_r.pdf}}
\newcommand{\Uidenm}{\includegraphics[scale=0.0043]{U_iden_m.pdf}}

\newcommand{\MPSU}{\includegraphics[scale=0.0182195]{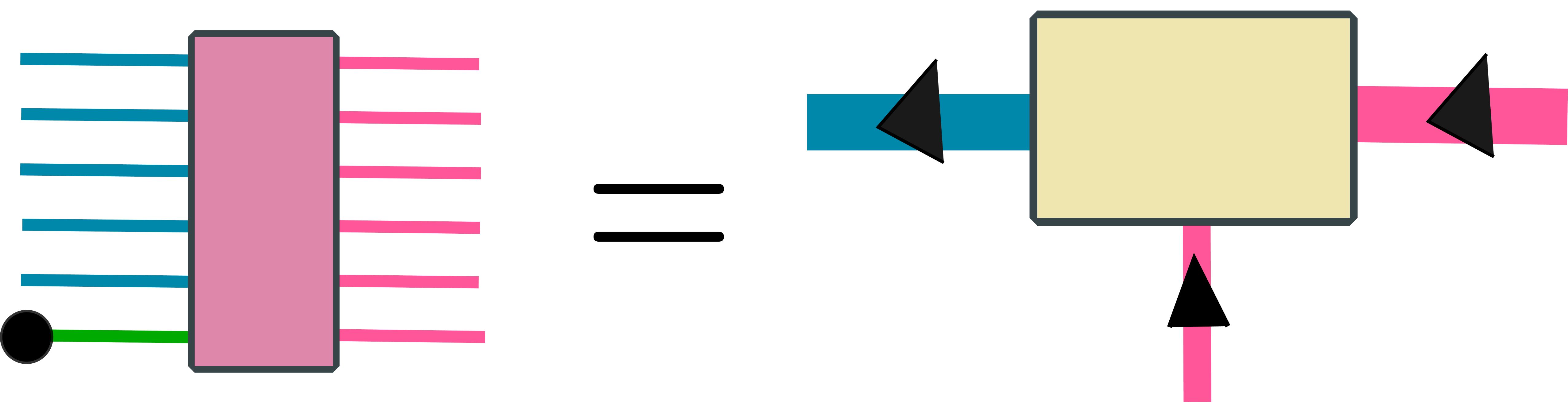}}
\newcommand{\Lines}{\includegraphics[scale=0.005, height=1em]{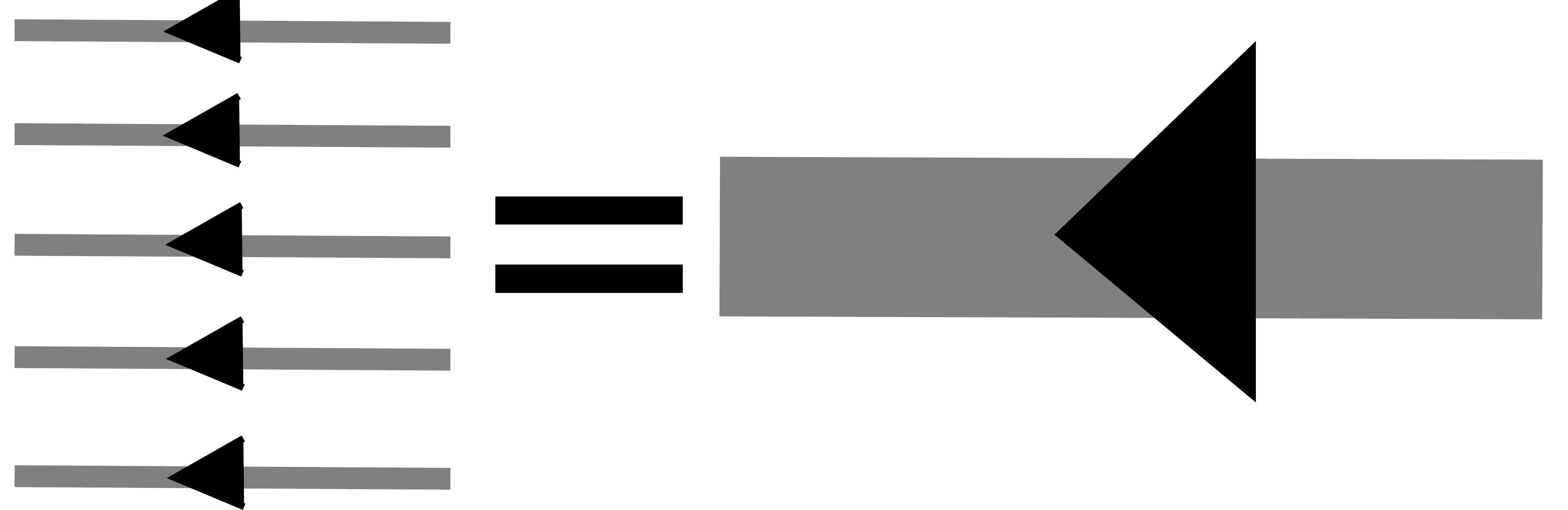}}

\newcommand{\LinesTWO}{\includegraphics[scale=0.005]{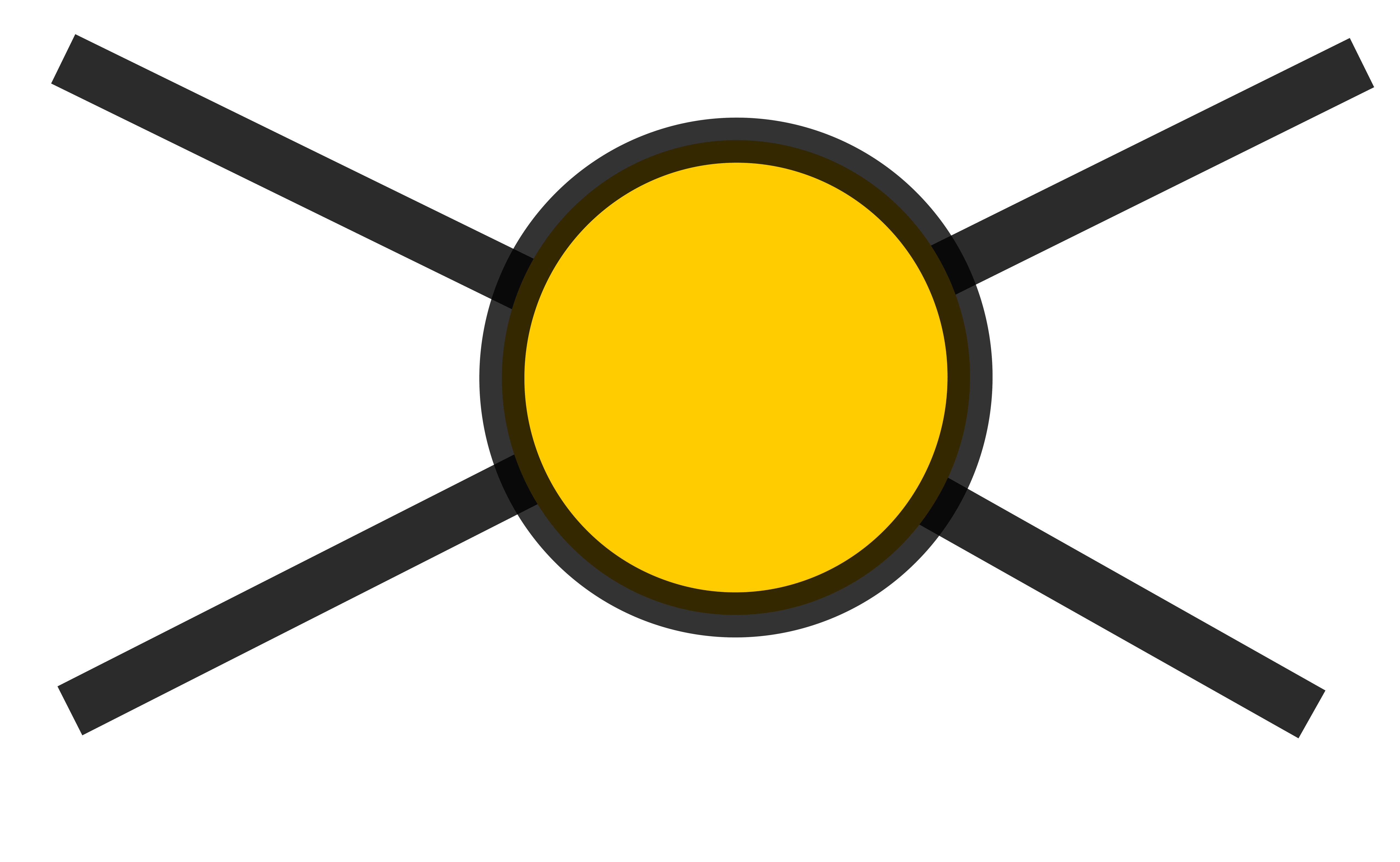}}
\newcommand{\pLinesTWO}{\includegraphics[scale=0.005]{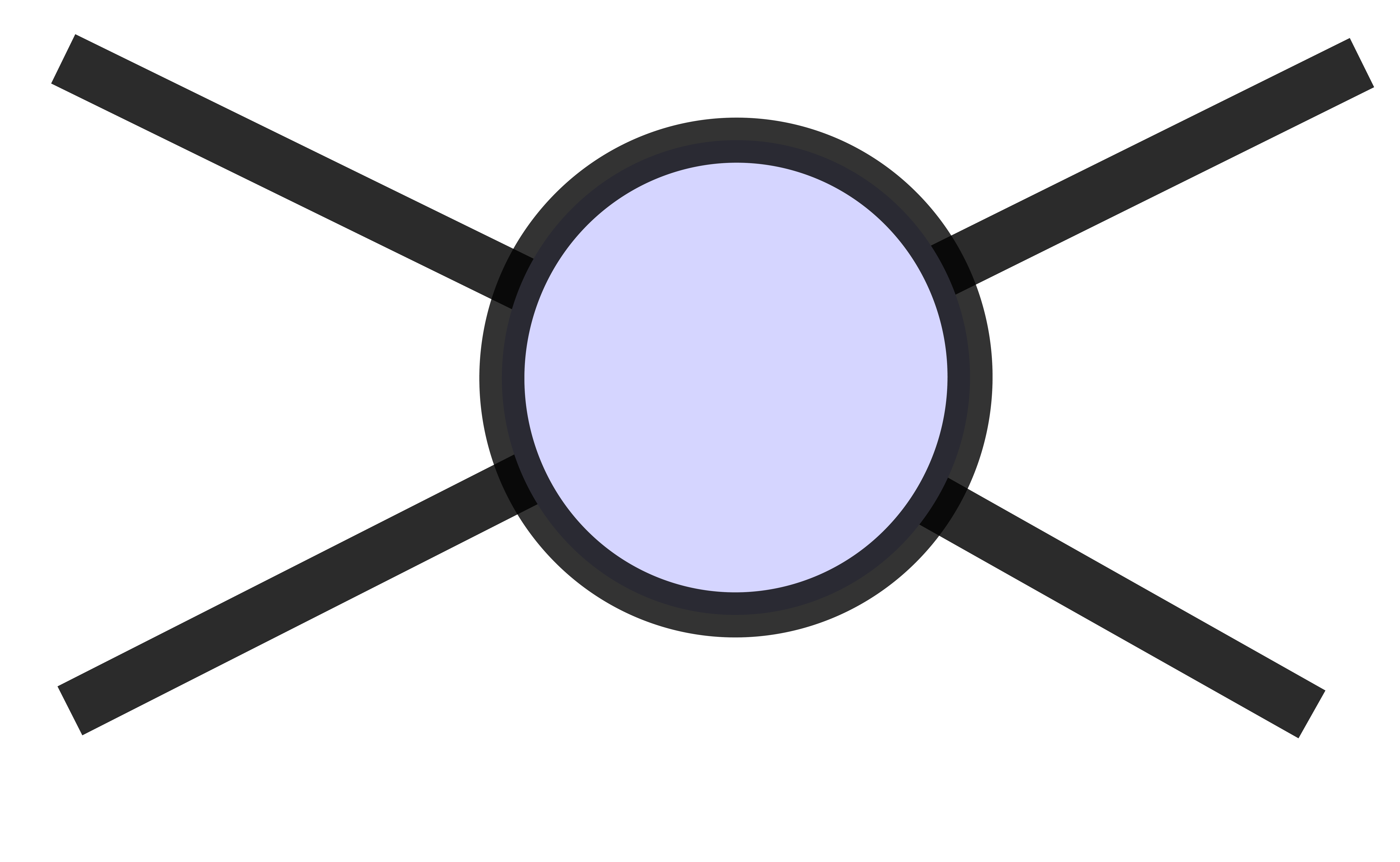}}

\newcommand{\disentangler}{\includegraphics[scale=0.007]{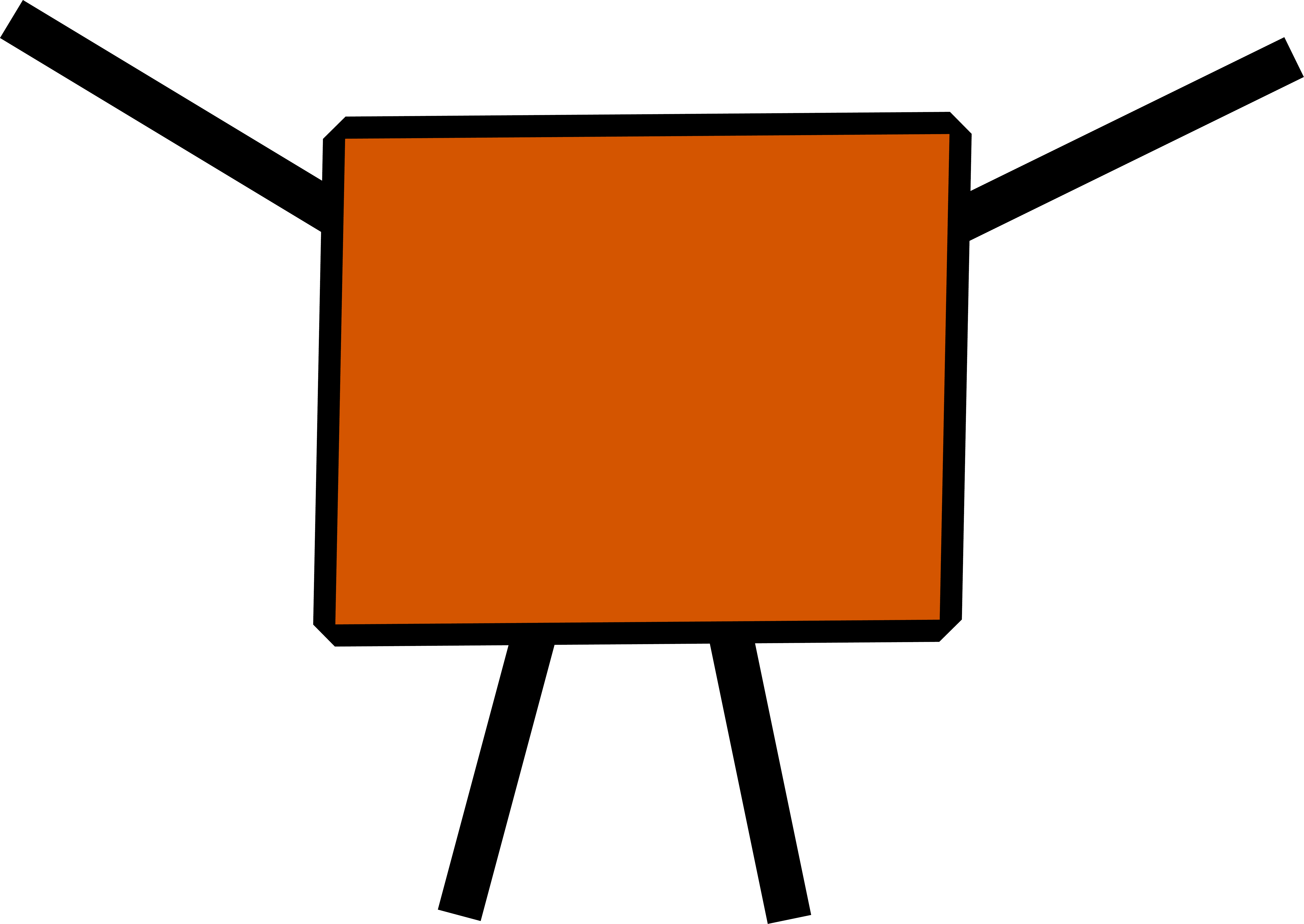}}

\begin{document}

\title{The Variational Power of Quantum Circuit Tensor Networks}
\author{Reza Haghshenas}
\email{haqshena@caltech.edu}
\affiliation{Division of Chemistry and Chemical Engineering, California Institute of Technology, Pasadena, California 91125, USA}

\author{Johnnie Gray}
\email{jgray@caltech.edu}
\affiliation{Division of Chemistry and Chemical Engineering, California Institute of Technology, Pasadena, California 91125, USA}

\author{Andrew C. Potter}
\affiliation{Department of Physics, University of Texas at Austin, Austin, TX 78712, USA}
\affiliation{Department of Physics and Astronomy, and Quantum Matter Institute,
University of British Columbia, Vancouver, BC, Canada V6T 1Z1}

\author{Garnet Kin-Lic Chan}
\email{garnetc@caltech.edu}
\affiliation{Division of Chemistry and Chemical Engineering, California Institute of Technology, Pasadena, California 91125, USA}

\begin{abstract}
We characterize the variational power of quantum circuit tensor networks in the representation of physical many-body ground-states. 
Such tensor networks are formed by replacing the dense block unitaries and isometries in standard tensor networks by local quantum circuits. We explore both quantum circuit matrix product states and the quantum circuit multi-scale entanglement renormalization ansatz, and introduce an adaptive method to optimize the resulting circuits to high fidelity with more than $10^4$ parameters. We benchmark their expressiveness against standard tensor networks, as well as other common circuit architectures, for the 1D/2D Heisenberg and 1D Fermi-Hubbard models.
We find  quantum circuit tensor networks to be substantially more expressive than other quantum circuits for these problems, and that they can even be more compact than standard tensor networks. Extrapolating to circuit depths which can no longer be emulated classically, this suggests a region of advantage in quantum expressiveness in the representation of physical ground-states.
\end{abstract}
\maketitle

\section{Introduction}

Advances in digital quantum computing have led to renewed interest in quantum circuit representations of many-body states. For this purpose, it is crucial to understand the representational power and trainability of different circuit architectures.
Out of the many possible architectures, one promising choice is circuits derived from tensor network states used in classical simulations of quantum states with limited entanglement. 
Because of the close connections between tensor networks and quantum circuits, and
the significant numerical experience using them in classical simulations,
they provide a natural setting to define the boundary between classical and quantum capabilities for quantum simulation. 
The simplest measure of the classical complexity of a tensor network is the tensor bond dimension. Consequently, one can construct a tensor network that is hard to simulate classically by 
devising a quantum circuit that maps to a tensor network with a very large bond dimension, in a small number of circuit elements. For example, one can construct quantum circuits that generate tensors with bond dimensions exponential in the circuit depth.
This is the idea behind ``deep'' or quantum circuit tensor networks which have been of
interest for different applications of quantum devices~\cite{kim:2017, Cong:2019, Huggins:2019, fossfeig:2020, Bauer:2020, Lubasch:2020, Barratt:2021}. Further, when combined with mid-circuit measurements and qubit reuse, some of these methods allow simulation of large-scale quantum systems with few physical qubits~\cite{fossfeig:2020,foss2021entanglement,chertkov2021holographic}.

However, constructing a class of states that is hard to represent classically is not sufficient to advance the simulation of physical systems. 
In the context of physical quantum many-body states, we must address additional questions (i)  are sparsely parameterized quantum ``circuit" tensor networks capable of representing physical states more efficiently than the ``dense" tensor networks (i.e. where all elements of the tensors can be independently varied) traditionally used in classical simulation? (ii) and for this task, what are the optimal circuit architectures and optimization protocols? 
The answers have potential implications not only for quantum simulations, but also for   classical simulations with tensor networks. For example 
an affirmative answer to (i) would suggest that 
it may sometimes be better to classically simulate with the quantum circuit tensor network directly, rather than via the classical dense tensor network, so long
as the circuit depth and tensor values support efficient classical contraction and/or approximation. 

Some work to answer questions (i) and (ii) has already appeared, such as in Refs~\cite{Uvarov:2020, Lin:2021,slattery:2021unitary, Haghshenas:2021}.  In this work, our focus will be on establishing the variational power of
quantum tensor networks to capture quantum many-body ground-states. This is an application where traditional dense tensor networks do well, and is thus in some sense the hardest test for quantum circuit tensor networks to pass. We will focus in particular on understanding the resources (e.g. number of variational parameters) required, optimization strategies, and influence of circuit architecture on the results. Because of the large number of numerical experiments required, we will mainly focus on one-dimensional quantum many-body states, although we present suggestive findings on two-dimensional problems also. As we shall demonstrate, with careful optimization strategies, quantum circuit tensor networks are very expressive, and in some cases, even more expressive than classical dense tensor networks. This suggests a regime where a quantum advantage in the sense of \textit{expressiveness} may be observed in physical ground-state simulations.

The structure of the paper is as follows. We first introduce quantum circuit tensor networks, review the mapping between two common tensor networks, the matrix product state and the multiscale entanglement renormalization ansatz\cite{Vidal:2008} (MERA), to block unitary circuits~\cite{Barratt:2021,fossfeig:2020}, and introduce different architectures for the local unitary representations.  We also provide some intuition regarding the expressiveness of different structures of quantum circuit ans\"{a}tze.
We then examine optimization strategies for such circuits. We next carry out a detailed comparison between the quantum circuit tensor networks, classical dense tensor networks, and reference circuit classes studied in the literature, to evaluate their variational efficiency and power, for both energies and correlation functions. We finish with a discussion of our findings in the context of computational quantum advantage.

\begin{figure}
\begin{center}
\includegraphics[width=1.0 \linewidth]{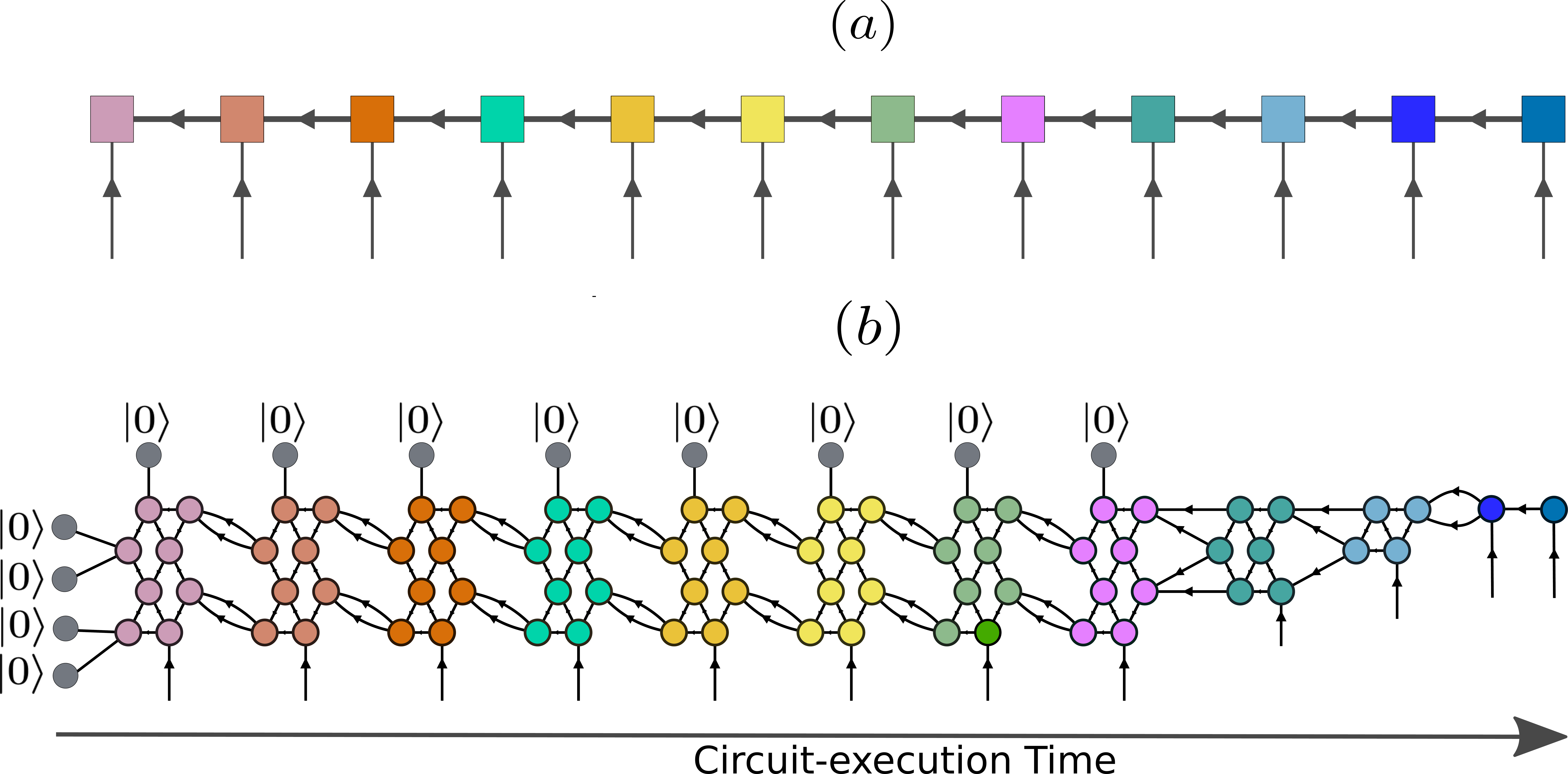} 
  \caption{(Color online) Schematic representation of a matrix product state and its quantum circuit in (a) right canonical form for a system with $L=12$ (or equivalently $12$ qubits). The square tensors 
 represent isometric tensors. (b) A quantum circuit MPS with four bond qubits ($q=4$) and a local brickwall circuit structure with depth $\tau=4$. The circle tensors 
are two-qubit unitary gates, while the black tensors denote register qubits initialized in the $|0\rangle$ state. Note that the arrows associated with the MPS and quantum circuit MPS follow the tensor network convention rather than the circuit convention, i.e. they are in the \emph{opposite} direction of circuit-execution time.}
  \label{fig:qmpsBW}
\end{center}
\end{figure}

\section{Tensor networks and quantum circuits}
\label{Sec:qmps}
\subsection{The Canonical Form of the Matrix Product State}

A matrix product state (MPS)\cite{Affleck:1987, Verstraete:2006, Verstraete:2008} is a tensor network consisting of a tensor for each site, connected by bonds in a one-dimensional geometry. It is represented diagrammatically by
\begin{equation} 
\label{EQ:mpsform}
\diagram{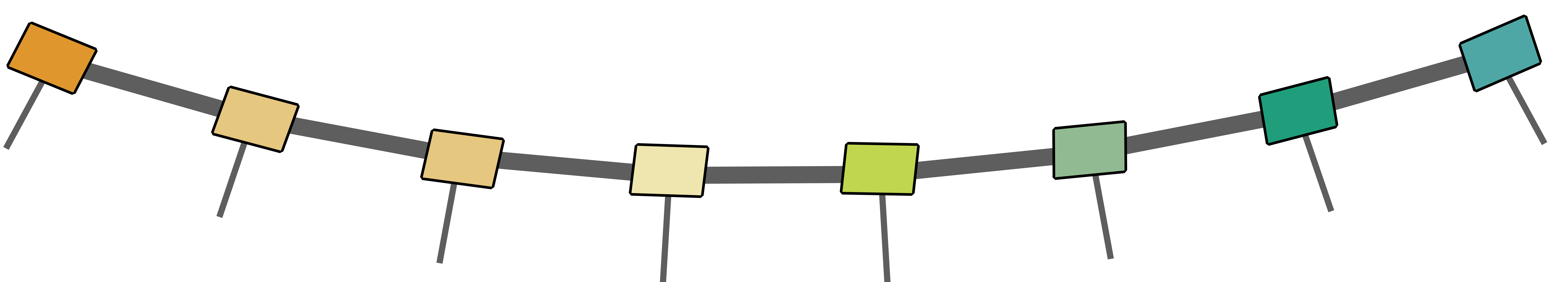}
\end{equation}
where each square tensor $\mpsten$ denotes a $D \times D \times d$-dimensional tensor. The open bonds index the $d$-dimensional physical Hilbert space of the site. (For example, if $d=2$, we can identify each site with a qubit).
The connected ``virtual'' bonds of dimension $D$ then control the number of parameters and thus the variational power of the MPS (or more physically, the maximum bipartite entanglement at each bipartition in the network). 

The individual tensors in an MPS are not in unique correspondence with a given quantum state due to gauge degrees of freedom: the state is invariant under insertion of a gauge matrix and its inverse $G, G^{-1}$ between two tensors (along a virtual bond). To fix the gauge degrees of freedom, a MPS can be recast into a canonical form\cite{Garcia:2007}. In canonical form, the tensors satisfy additional isometric or normalization constraints, but for a (normalized) MPS, this imposes no loss of representational power. 
A simple algorithm to obtain the canonical form is to perform a sequence of $QR$ ($LQ$) decompositions of the tensors; doing this from left to right (right to left) brings an arbitrary MPS into left (right) canonical form\cite{Vidal:2003OCT}. For example, the  right canonical form of the MPS in Eq.~\ref{EQ:mpsform} can be represented  by the following diagram:
\begin{equation*} 
\label{EQ:SplittingTR}
\diagram{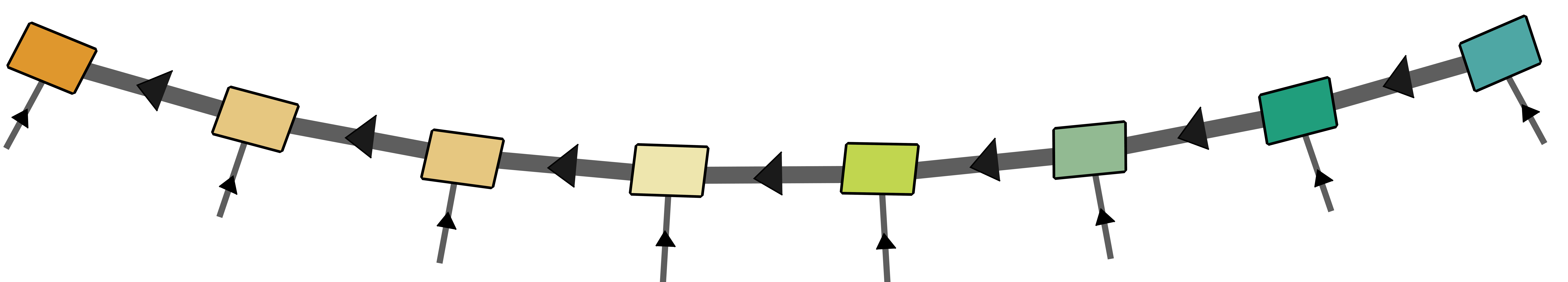}
\end{equation*}
where every square tensor with arrows $\Uten$ is an isometric tensor.
 The isometric condition is satisfied for contractions performed on the incoming bonds $\Uiden$, where the left side of the equality is the identity matrix. The isometric constraint fixes the gauge freedom up to permutations of bond indices. 
 In a similar way, the right canonical form is defined by tensors which satisfy an isometric condition $\Uidenr$, while a ``mixed canonical form" (central to the density matrix renormalization group (DMRG)~\cite{White:1992}) is obtained by combining left and right canonical forms around a given central site, with the central tensor satisfying the condition $\Uidenm$. When the tensors in a canonical MPS satisfy isometric conditions, the MPS is an example of an ``isometric" tensor network~\cite{Haghshenas:2019, Zaletel:2020}. The number of independent variational parameters in a canonical MPS of length $L$, with all elements real, scales asymptotically as $\sim L \times D(3D-1)/2$. We will refer to these standard MPS
 as ``dense'' MPS (dMPS), since the tensors assume their most flexible parametrization, in contrast to the ``sparse'' parametrization by a quantum circuit used later. Because of the close relationship between the standard MPS formulation and the DMRG, we will sometimes use the term DMRG.

\subsection{Quantum circuit MPS}
\label{subSec:qmps}
Given the canonical form of the MPS, the mapping to a block unitary quantum circuit follows a simple procedure~\cite{cirac:2007,Barratt:2021,fossfeig:2020}.
This is because isometric tensors can be viewed as arising from the application of a unitary tensor to a partial set of inputs. Concretely, the steps are as follows: $(i)$ The virtual bonds of dimension $D$ (thick lines) are viewed as $q$ qubit bonds (aggregate dimension $D=2^q$) (thin lines). Graphically, this is the relation $\Lines$. (Note that, following standard MPS conventions, the arrows on these diagrams indicate right-canonical form, and are opposite to the execution-time direction in the associated quantum circuit).
$(ii)$ The isometric tensors are viewed as columns of a block unitary matrix with one fixed output qubit (denoted by a black dot, here assuming $d=2$),  i.e. $\MPSU$. This mapping generates the MPS via a block unitary circuit, where each block unitary is a matrix of dimension $dD\times dD$.

In the above mapping, the variational space of normalized states generated by the block unitary circuit and the standard ``dense MPS" (dMPS) are exactly the same. However, we can imagine replacing the block unitary by a local circuit of two-qubit gates of given depth. One can then create a block unitary that acts on a set of $q$ qubits, using as few as $O(q)$ two-qubit gates.  
This allows us to generate an entangled state that lives in the variational space of a $D=2^q$ dMPS, with as few as $O(q)$ variational parameters per site.
We refer to an MPS where the block unitary
is encoded as a local circuit as a ``quantum circuit MPS'' (qMPS).

\begin{figure}
\begin{center}
\includegraphics[width=1.0 \linewidth]{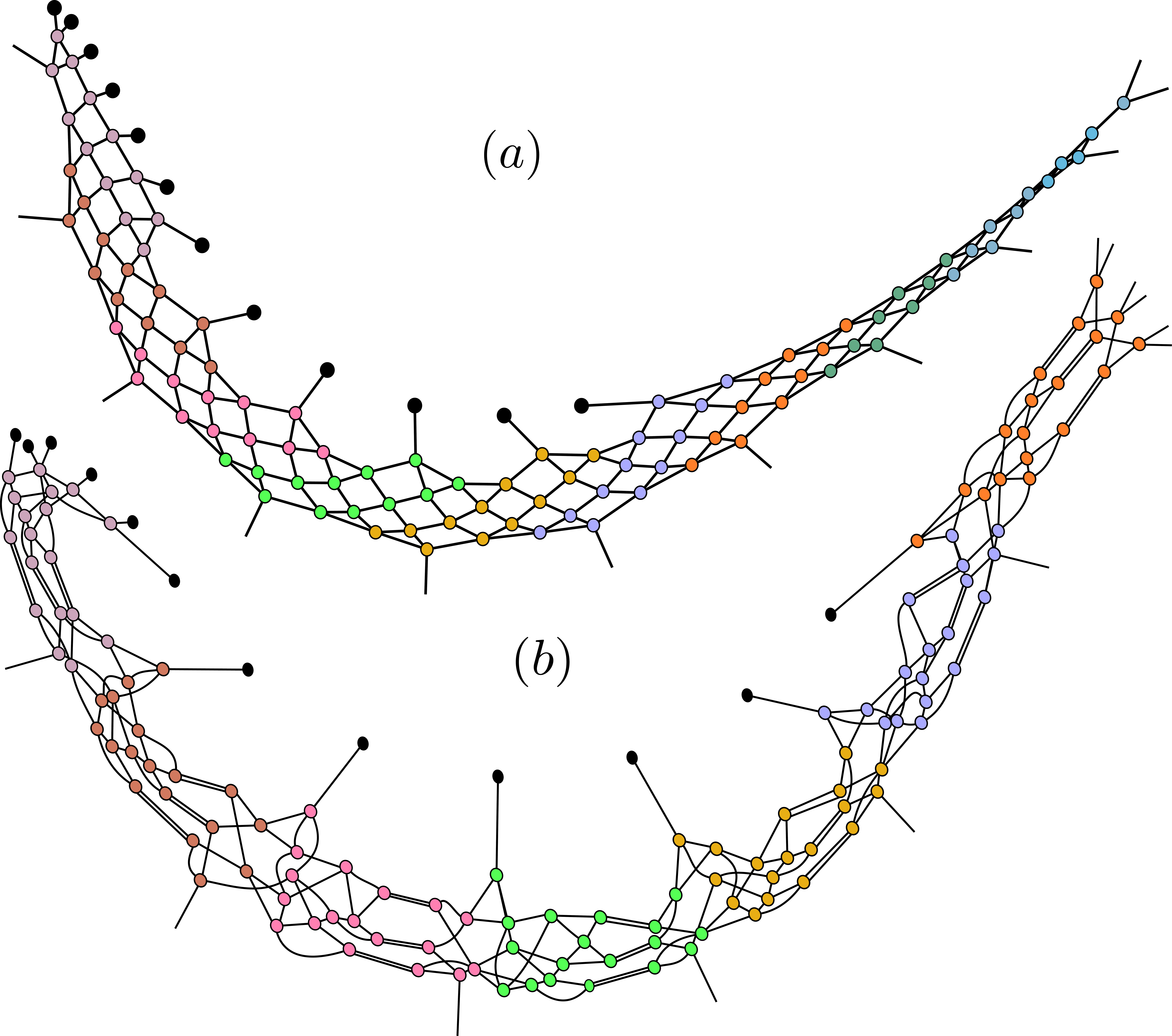} 
  \caption{(Color online) Schematic representation of a qMPS circuit with  (a) ladder and (b) MERA internal local circuits, with local depth $\tau =2, 4$ and bond qubits $q =5, 6$, respectively.}
  \label{fig:qMPSqMERA}
\end{center}
\end{figure}

There is much freedom to choose the structure of the local circuit. 
Here, we explore several local circuit structures:
(i) a brick-wall circuit, denoted graphically as
\begin{equation*} 
\diagrammB{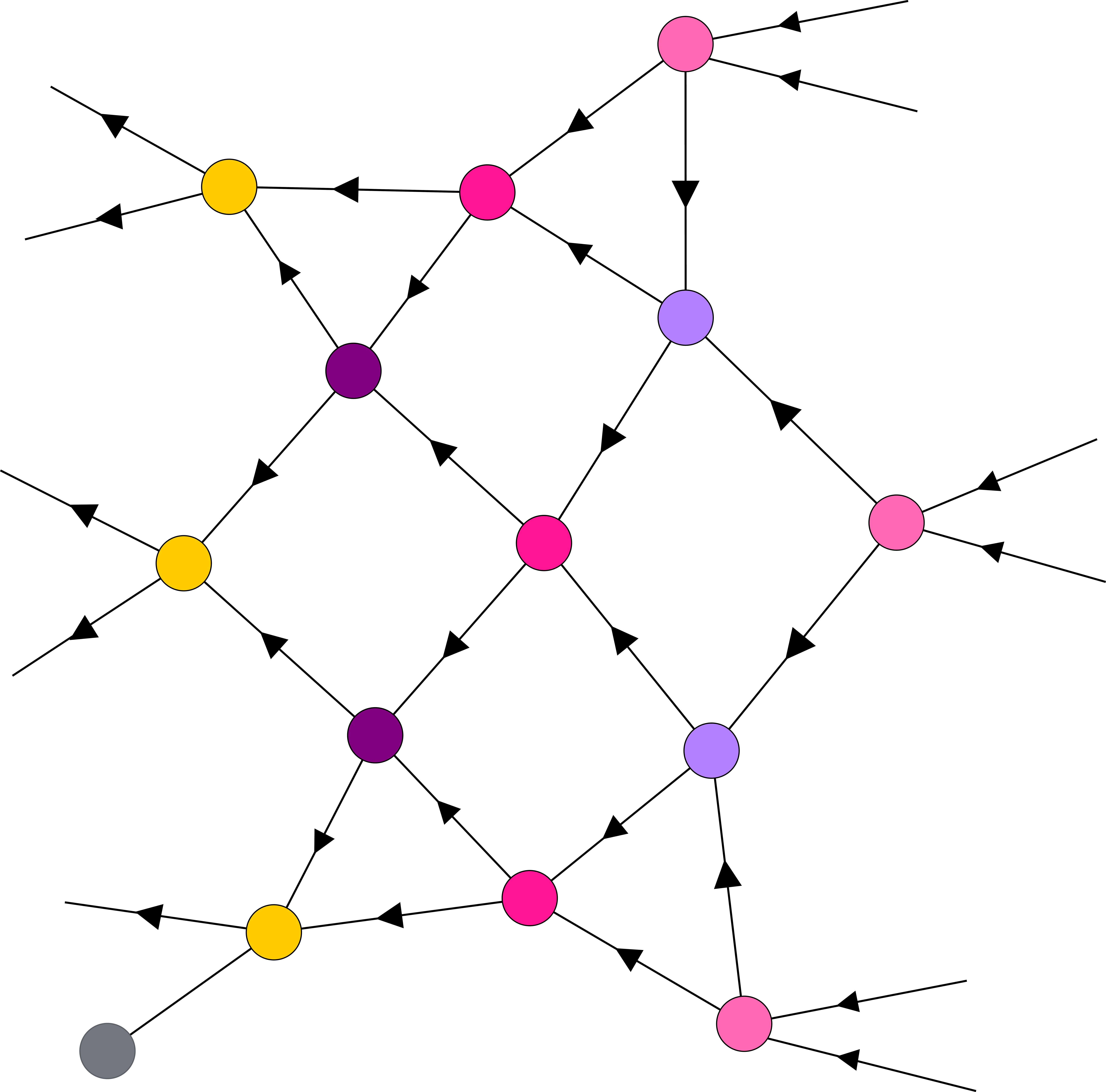}
\end{equation*}
with densely connected nearest-neighbour two-body
unitary gates (circle tensors $\LinesTWO$). Throughout, colors visually distinguish different circuit layers, but each gate of a given color implements a distinct gate. 
For the brick-wall circuit, we refer to a layer of even gates and a layer of odd gates  as two layers, thus the above circuit has a circuit depth of $\tau=6$. 
In a brick-wall circuit, correlations spread slowly with increasing $\tau$, as the effective correlation length $\xi$ grows linearly with circuit depth $\xi \sim \tau$. (ii) A ladder circuit,
 for example,
 \begin{equation*} 
\diagrammZ{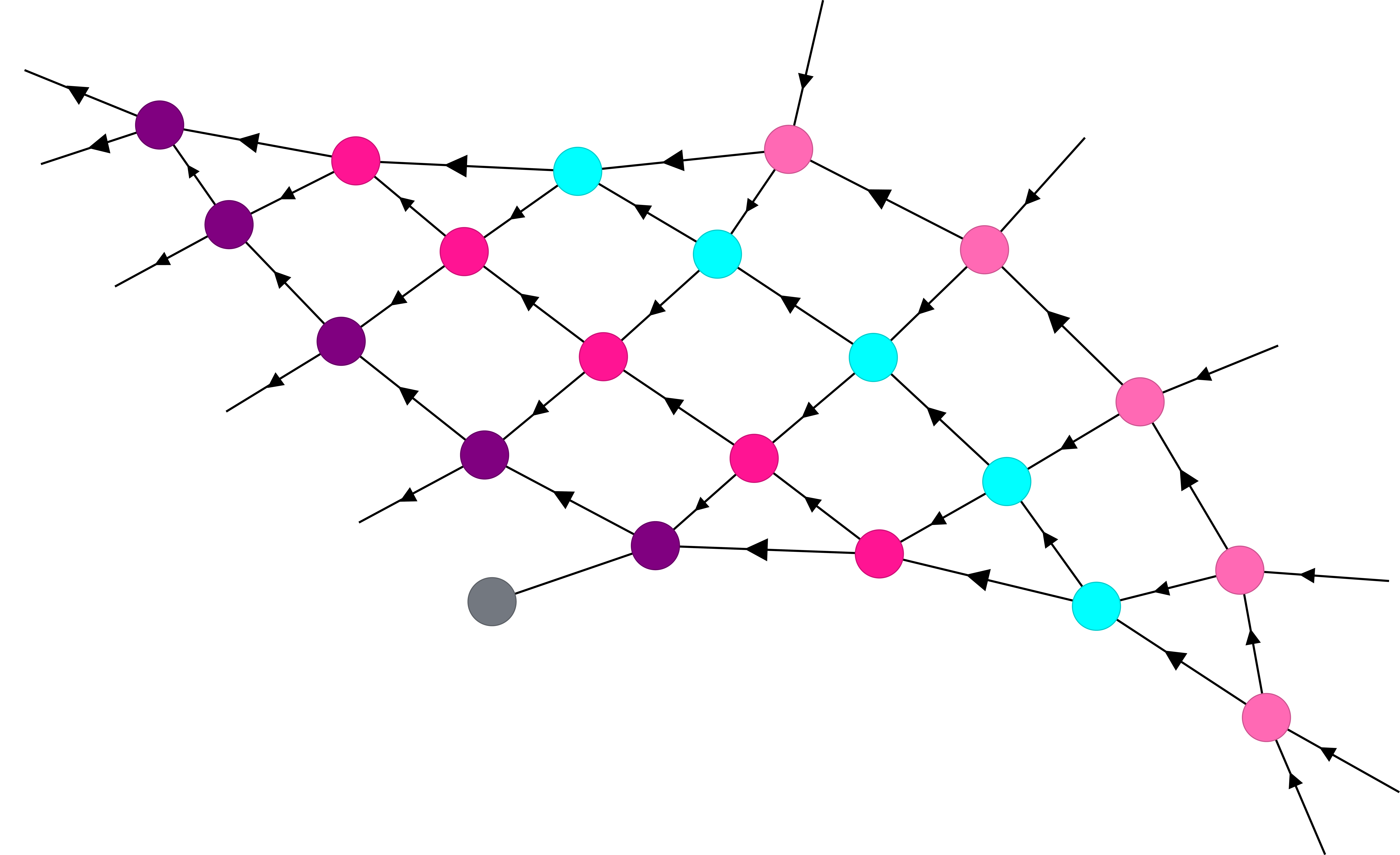}
\end{equation*}
where we label the circuit above as depth $\tau=4$. Correlations propagate more efficiently in this structure: with $\tau=1$, the first and last qubits are already correlated. (iii) A MERA structure
\begin{equation*} 
\diagrammM{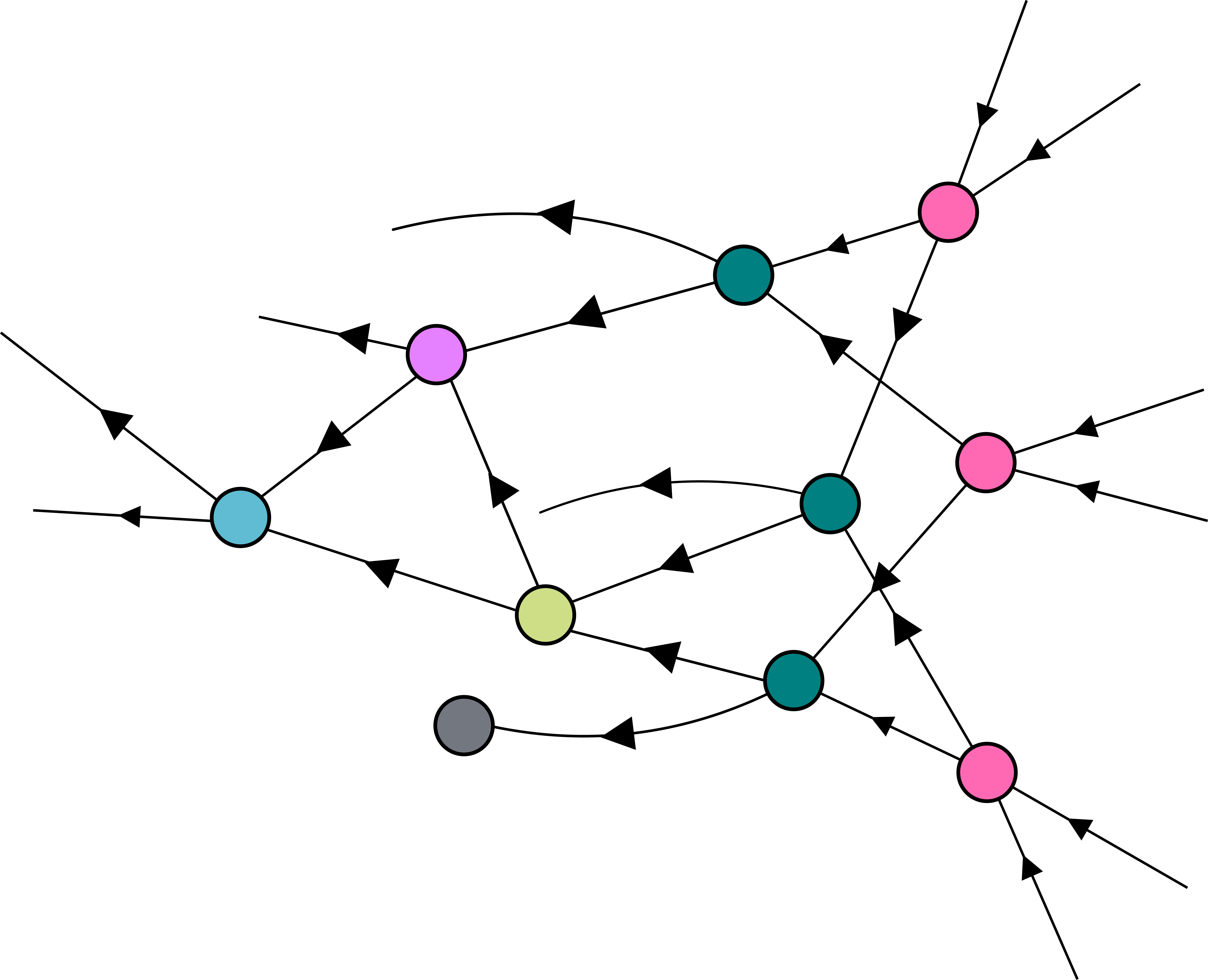}
\end{equation*}
which represents a binary MERA\cite{Evenbly:2009} with depth $\tau=5$. Properties of MERA circuits in general are discussed more in the section below. Note that here, however, the MERA structure is only being used for the local circuit (i.e. a MERA quantum circuit, rather than a quantum circuit MERA) and globally, the ansatz is still a qMPS.
An example of the final circuit structure of the qMPS using a local brick-wall circuit is shown in Fig.~\ref{fig:qmpsBW}. Corresponding figures for qMPS with local ladder and MERA circuits are shown in Fig.~\ref{fig:qMPSqMERA}.

Overall, the variational power of the qMPS ansatz is determined by three factors: (i) number of qubits $q$ on which each block unitary circuit acts (placing the qMPS in the variational space of a dense MPS with $D=2^q$), (ii) the number of gates in the local circuit, (iii) the internal structure of the local circuit. Note that the number of gates in each layer differs between the local circuit structures, thus $\tau$ should not be directly compared between the different structures. Instead, the number of gates (or equivalently, variational parameters)  asymptotically behaves as  $\sim \frac{1}{2}\tau L (q+1)$, $\sim \tau L (q+1) $ and $\sim 2 \tau L (q+1) $ for the brick-wall, ladder, and MERA structures, respectively. In the numerical simulations, we will refer to these three kinds of qMPS circuits as qMPS-b, qMPS-l, and qMPS-m, respectively.

\begin{figure}
\begin{center}
\includegraphics[width=0.80 \linewidth]{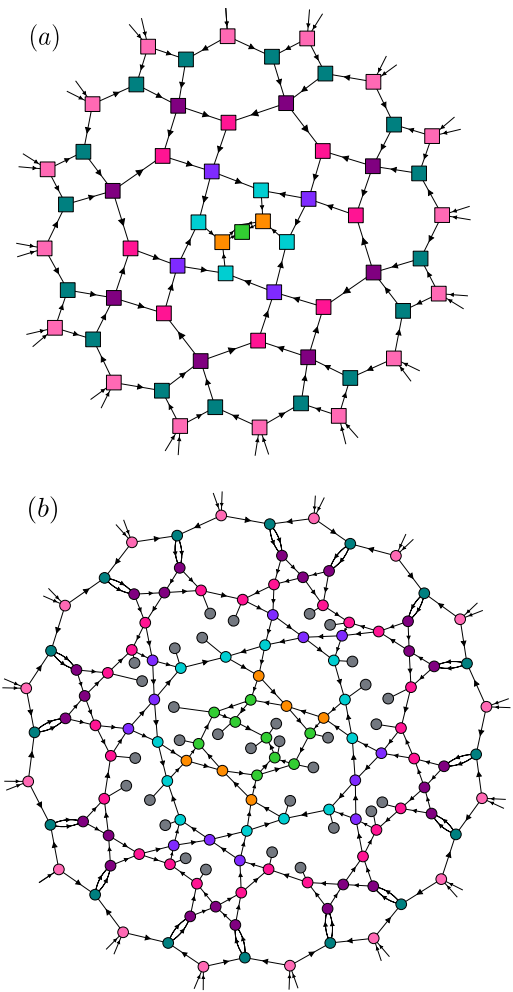} 
  \caption{(Color online) (a) A binary MERA wave function for $L=32$ qubits, and (b) its quantum circuit, i.e. qMERA, where the local quantum circuit is chosen to have a brick-wall structure with $q=2$ and $\tau=2$. The different layers of the MERA are shown in different colors.}
  \label{fig:qmera}
\end{center}
\end{figure}

\subsection{Quantum circuit MERA}

Another commonly used tensor network is the MERA. This
 is a  tensor network state where the tensors are arranged to introduce bipartite entanglement at multiple scales. In this ansatz, isometric tensors perform  coarse-graining while unitary disentanglers are applied to remove short-range entanglement at the different length scales. We show a binary form of MERA in Fig.~\ref{fig:qmera}(a) with unitary disentanglers and isometric coarse-graining tensors $\disentangler$ distinguished by different colors in each layer. We will refer to the standard form of MERA where all tensors are assumed dense as dense MERA (dMERA).

 Since MERAs are isometric tensor networks by construction, like in the MPS canonical form, a quantum circuit MERA (i.e. a circuit whose global structure is derived from the MERA) i.e. qMERA can be straightforwardly obtained by decomposing both the block isometric and unitary tensors into local circuits with a finite depth $\tau$ and given internal structure as already discussed. A graphical illustration of a qMERA is shown in Fig.~\ref{fig:qmera}(b). An important difference between a qMERA and a qMPS is the presence of a structured set of long-range unitary gates.
In 1D, this allows the qMERA to capture critical entanglement beyond the area law with only a polynomial number of gates~\cite{Evenbly:2011}. 

Similarly to the qMPS, the variational power of the qMERA ansatz is determined by three factors, i.e. $q$, the number of gates, and the internal structure of local circuits. In this work, we only consider qMERA with local brick-wall circuits as depicted in Fig.~\ref{fig:qmera}(b)\cite{mera:note}, which we will refer to as qMERA-b in the numerical studies.

\begin{figure}
\begin{center}
\includegraphics[width=1.0 \linewidth]{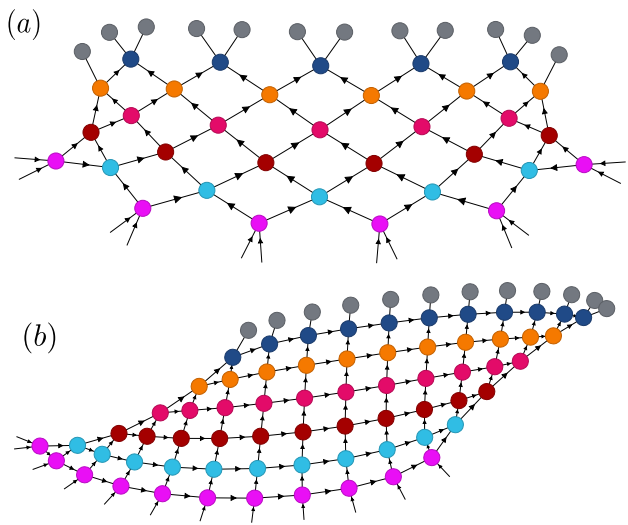} 
  \caption{(Color online) Global quantum circuits with (a) brick-wall and (b) ladder structures. Execution time flows from top to bottom. }
  \label{fig:circuit}
\end{center}
\end{figure}

\begin{figure}
\begin{center}
\includegraphics[width=.90 \linewidth]{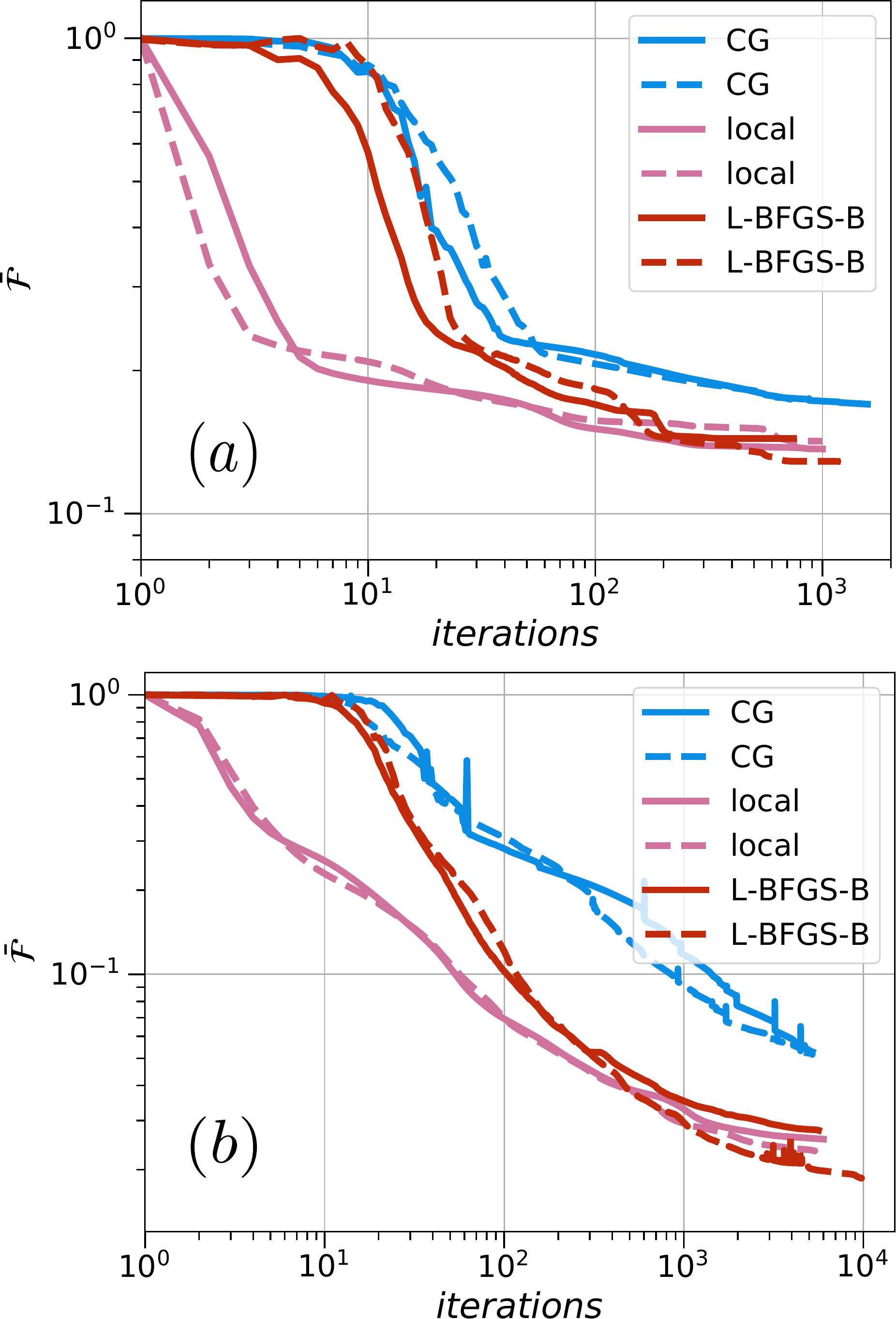} 
  \caption{(Color online) The influence of the choice of optimization method. (a) The infidelity $\bar{\mathcal{F}}$
  versus iteration number for the global brick-wall ansatz QC-b with $L=16, \tau=6$, optimized by CG, L-BFGS-B and local methods. The solid and dashed lines represent two different initial starting states, chosen from a uniform random distribution for the tensors. (b) The same plot for a qMPS-b with $L=24, q=4,\tau=4$. The targeted wave function is the ground state of Heisenberg model $\mathcal{H}_\text{H}$.}
  \label{fig:fidelOpt}
\end{center}
\end{figure}

\begin{figure}
\begin{center}
\includegraphics[width=1.0 \linewidth]{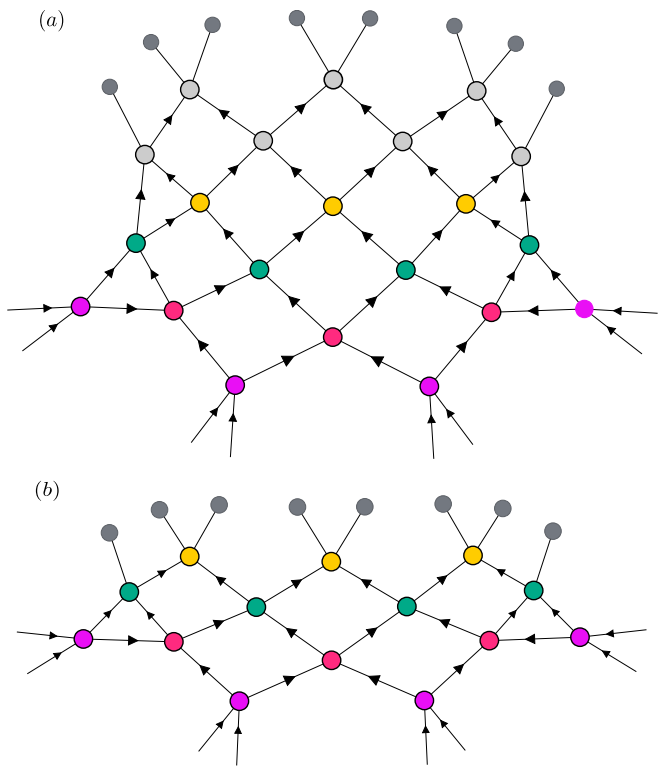} 
  \caption{(Color online) Adaptive initialization of a QC-b circuit with (a) $\tau=6$ from an optimized smaller circuit (b) with $\tau=4$. The gray gates 
denote identity operators with some small random perturbations. Note that execution time flows from top to bottom (opposite to the arrow direction).}
  \label{fig:init}
\end{center}
\end{figure}

\subsection{Global quantum circuit ansatz}

To place the performance of the quantum circuit tensor networks in context, we also consider global brick-wall and ladder circuit ans\"atze, as depicted in
Fig.~\ref{fig:circuit}. These are referred to as QC-b and QC-l in the numerical studies below.

\subsection{Properties of different quantum circuit ansatz}

\label{sec:analysis}

All the above ansatz are universal approximators in the sense that with sufficient numbers of parameters (for example, controlled by $q$ and $\tau$ in the quantum circuit tensor networks) they can represent any state. Certain types of ground-state might be more compactly represented by one ansatz than another, but it is difficult to make general statements without numerical studies, as performed below. However, here we briefly provide some intuition into the relationship between circuit structure and expressivity, and the connection between the different ansatze.

\begin{figure}
\begin{center}
\includegraphics[width=1.0 \linewidth]{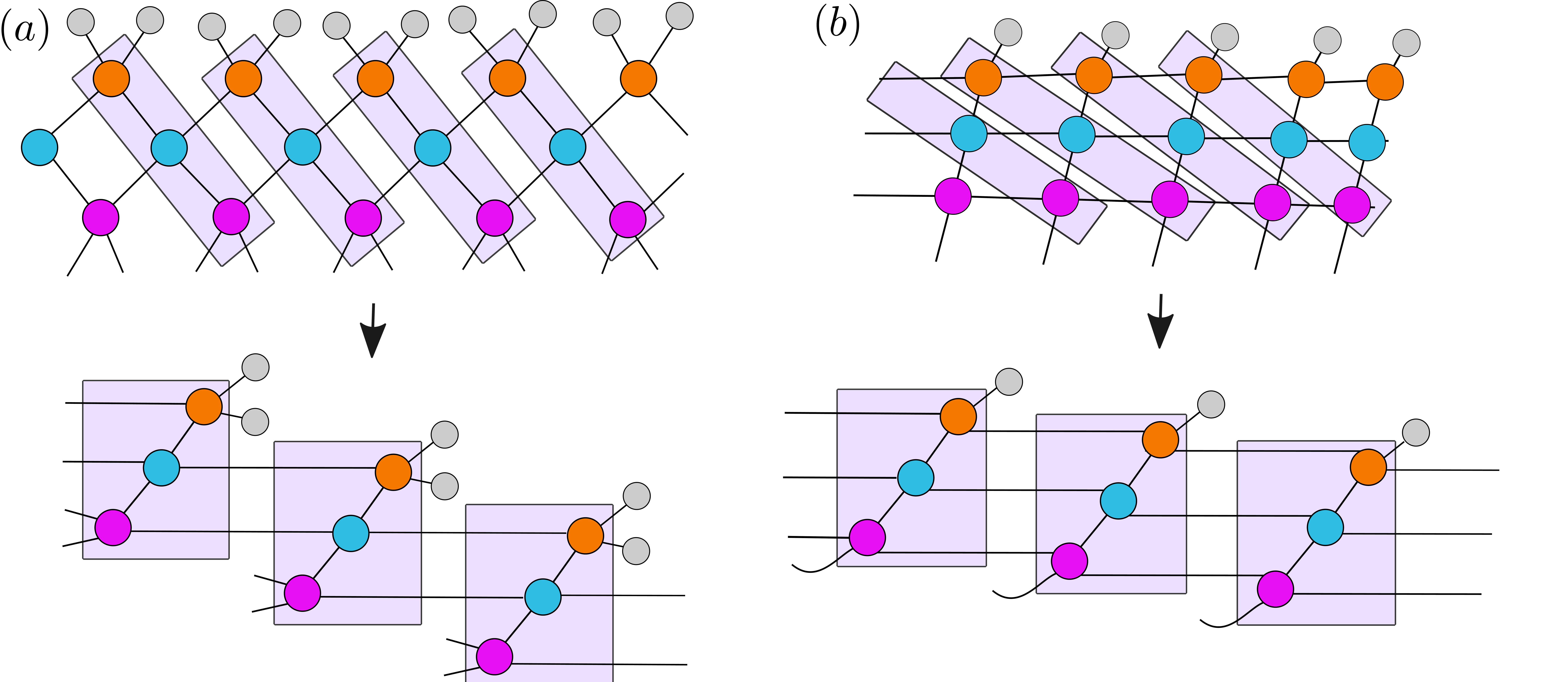} 
  \caption{(Color online) Mapping from global quantum circuit ansatz to qMPS for the (a) brick-wall and (b) ladder structures. Grouping the gates in the global ansatz (top) defines block-unitaries in the qMPS (bottom). Execution time flows from top to bottom.}
  \label{fig:qMPStoGC}
\end{center}
\end{figure}

We start with the relationship between the global quantum circuits and quantum MPS. Both the global brick-wall and ladder circuits can be directly transcribed into qMPS by grouping gates into block unitaries (Fig.~\ref{fig:qMPStoGC}). QC-b and QC-l circuits with $\tau$ layers map to qMPS with $q=\tau-1$ and $q=\tau$ block unitaries, respectively. Each block unitary has the sparsest possible entangled parametrization with  $\tau$ gates arranged in a ladder structure, with the ladder ascending in the opposite direction to the ascending direction of the qMPS block unitaries. In fact, the only difference between the global brick-wall and global ladder circuits when viewed from their corresponding qMPS, is that the QC-b block unitaries overlap on only $q-1$ qubits rather than the usual $q$ qubits. This choice of non-maximal overlap is what gives rise to the specific brick-wall lightcone, where correlations cannot  spread as quickly as in a general qMPS or a global ladder circuit.

Mapping QC-b and QC-l to qMPS reveals that the circuit structures prioritize reaching block unitaries of large size $q$ (large MPS bond dimension $D=2^q$) at the expense of accurately representing each unitary, as each block unitary is only minimally connected. If we assume, as seems likely, that the part of the dense MPS variational space of bond dimension $D=2^q$ required to represent a large variety of quantum ground-states of physical interest 
is not fully captured by these minimal local circuits, then QC-b and QC-l do not efficiently cover the variational space. One can see the influence of the block unitary circuit depth most dramatically in the expectation values of operators acting on the leftmost site. Because of the circuit ordering, such expectation values depend \textit{only} on the parametrization of the first block unitary. In the QC-b and \mbox{QC-l} ans\"atze one must increase the \textit{global} circuit depth in order to improve the leftmost local expectation values.

In the more general form of the qMPS, $\tau$ and $q$ can be independently varied. Thus qMPS is a superset of \mbox{QC-b} or \mbox{QC-l} and is more expressive, although the balance between $\tau$ and $q$ will be problem specific. The question is then whether the local circuit depth can be significantly reduced from that required to exactly parametrize a block unitary over $q$ qubits, which is exponential in $q$. There is room for optimism, as there are other powerful  variational states which map to sparse parametrizations of dense MPS. For example, correlator product states~\cite{changlani2009approximating,neuscamman2011nonstochastic,clark2018unifying,al2011capturing}, entangled plaquette states~\cite{mezzacapo2009ground,thibaut2019long}, and neural network quantum states~\cite{glasser2018neural} can be viewed as variational states parametrized by non-unitary gates. These can map to dense MPS with large $D$ (for example, capturing volume law scaling of entanglement with a polynomial number of variational parameters). However, numerical studies have shown that the number of variational parameters required in these ansatz for physical ground state problems can be fewer than in a dense MPS~\cite{changlani2009approximating}. 


Many of the above points also apply to the qMERA, in particular, the potential for sparse circuit representation of the unitaries and isometries arising in the dense MERA. In addition, the special geometric structure of the MERA  means that in 1D it spreads correlations while capturing logarithmic corrections to the entanglement law. This is an important formal distinction from the other quantum circuit structures considered here, although its importance for capturing the energies and correlation functions of finite systems in numerical studies must be established empirically.

\begin{figure}
\begin{center}
\includegraphics[width=.90 \linewidth]{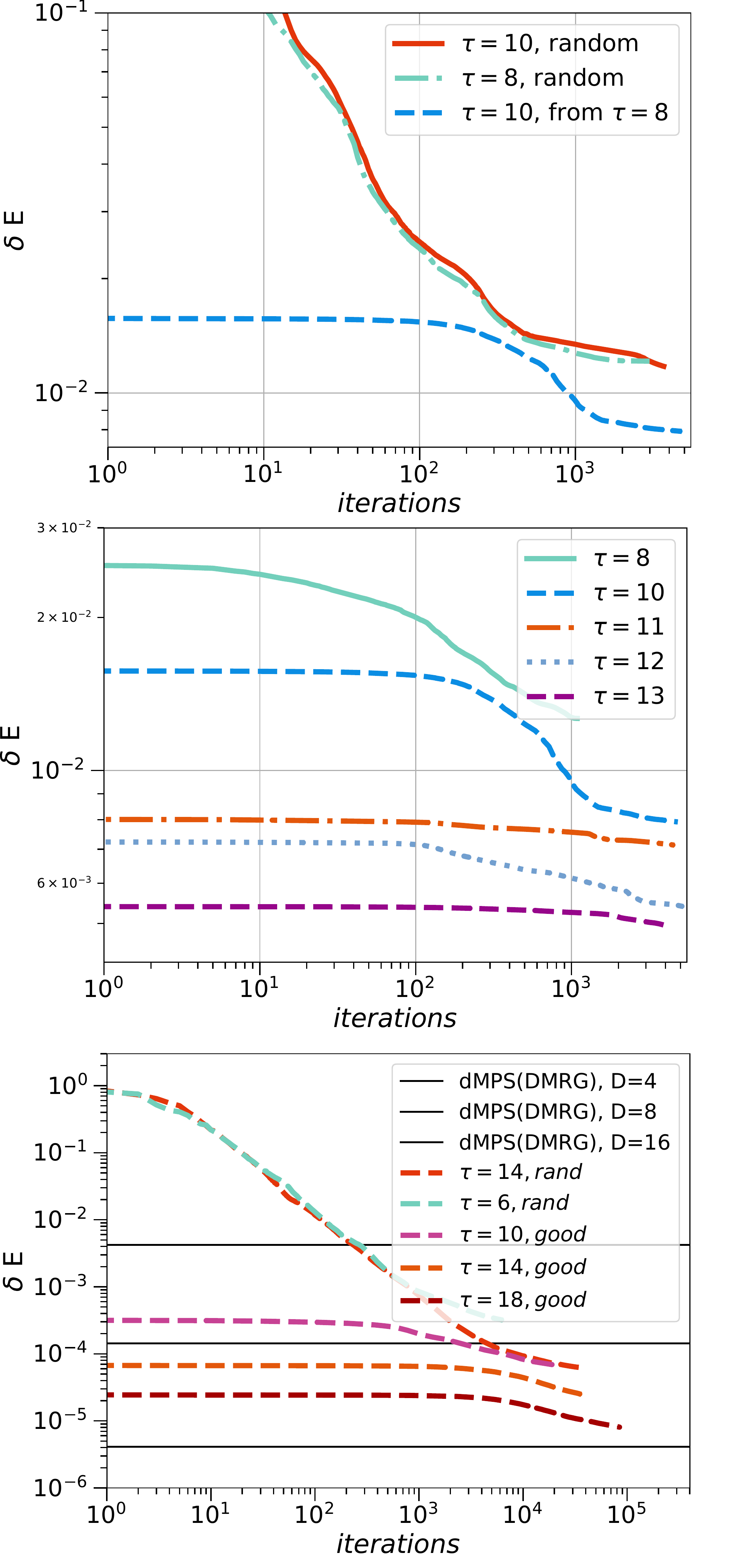} 
  \caption{(Color online) The role of the initial guess in the optimization. (a) The relative error of the ground-state energy $\delta E$ (of Heisenberg model $\mathcal{H}_\text{H}$) as a function of iteration number for a global brickwall quantum circuit (QC-b) with depth $\tau$. When the $\tau=10$ ansatz is initialized from the optimized $\tau=8$ parameters, we obtain a better minimum than from a random initial guess. Notice that when using a random initial guess the ansatz optimization can get stuck in a poor minimum, as seen by the $\tau=8$ (random) and $\tau=10$ (random) results, which obtain the same minimum. (b) A better initialization procedure using optimized circuit parameters from smaller depths guarantees that the relative error decreases monotonically when increasing circuit depth $\tau$. (c) Similar plot for a qMPS ansatz with $q=4$. We similarly find the relative error drops monotonically when increasing $\tau$ using initial guesses from a smaller $\tau$ ansatz (good), while initializing from random guesses (rand) results in optimizations which terminate at poor minima. Reference data from dense MPS (DMRG) also shown.  }
  \label{fig:initrole}
\end{center}
\end{figure}

\begin{figure*}[ht!]
\centering
\includegraphics[width=1.0 \linewidth]{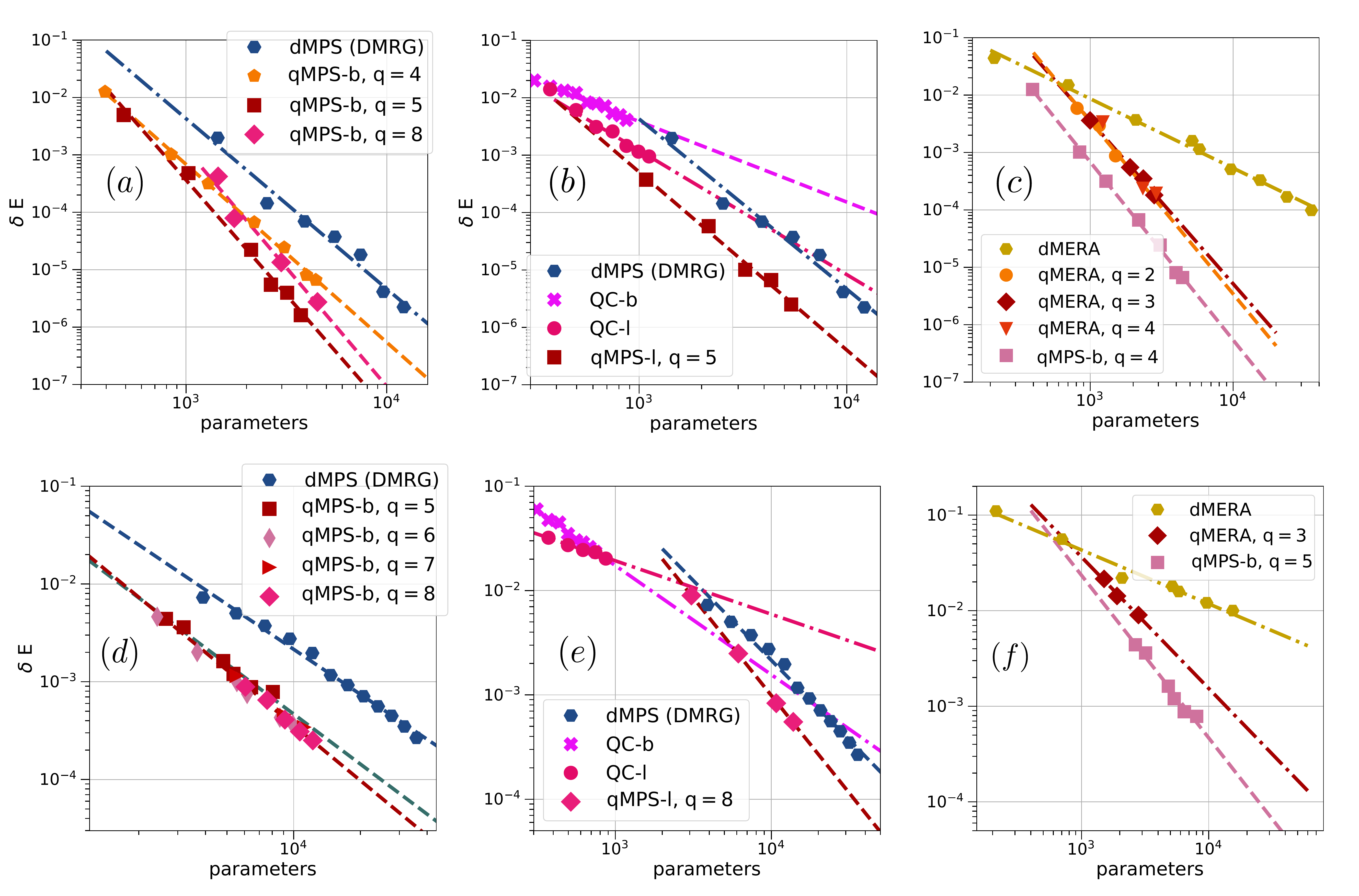}
\caption{ 
(Color online) Comparing the variational power of quantum circuit tensor networks (qMPS, qMERA), dense tensor networks (dMPS and dMERA), and global quantum circuits (QC). We show the relative energy error $\delta E$ versus the number of variational parameters in the ansatz for the Heisenberg (a,b,c) and Fermi-Hubbard (d, e, f)  models with $L=32$.  Indices 'b' and 'l' stand for brick-wall and ladder. (a, b) Comparison between qMPS with local brick-wall and ladder circuits with DMRG and QC with brick-wall and ladder structures.  The largest circuit depth used for  QC-b(l) and  qMPS-b(l) with $q=5$ is $\tau=14(9)$. (c) The performance of qMERA-b versus qMPS-b and dMERA. The largest parameter sets used for the qMERA-b ansatz correspond to $q=4, \tau=8$.  (d, e, f)  The same comparison for the Fermi-Hubbard model. The largest circuit depths for qMPS-b with $q=8$ and qMERA-b with $q=3$ are $\tau=32$ and $\tau=12$, respectively.}
\label{fig:deltaE}
\end{figure*}

\section{Numerical optimization studies}

\subsection{Algorithms}

The first question to answer in a numerical assessment of the variational power of an ansatz is how to optimize it. In this section, we investigate how to optimize the quantum circuit tensor network and global circuit ansatz considered in this work, by optimizing the parameters of the two-body unitary gates 
$\LinesTWO$. We assume each two-body unitary is a general $SO(4)$ unitary (i.e. real-valued unitary) with a 6 dimensional variational space\cite{counting:note}. We use two optimization algorithms: (i) a local ``DMRG''-like optimization, where we sweep through the unitary gates $\LinesTWO$, optimizing them one at a time while holding the others fixed; (ii) a global gradient-based optimization, where all variational parameters are updated at the same time. In the local optimization scheme, a linearization of the problem, similar to the one used in MERA optimizations~\cite{Evenbly:2009}, is used to find locally optimal gates. In the global gradient-based scheme, the global gradient (i.e. the first-order derivative with respect to all variational parameters $ \partial (\cdots,\pLinesTWO, \LinesTWO, \cdots)$) is analytically calculated by automatic differentiation as supported in \texttt{quimb}\cite{gray2018quimb}. 
The energy is computed via exact contraction\cite{gray2021hyper} of either a matrix product operator representation of the Hamiltonian (MPS) or a sum of local terms (MERA).
The unitary constraints are enforced by differentiating through a function that maps the gate parameters to a unitary matrix. The cost-function minimization is then performed using either the Conjugate Gradient (CG) or Limited-Memory Broyden–Fletcher–Goldfarb–Shanno  (L-BFGS-B) algorithms~\cite{nocedal1980updating,Jorge:2006}. The algorithms are stopped once the relative change in energy is less than $10^{-8}$.

\subsection{Model Hamiltonians}
We choose to study the 1D Heisenberg and Fermi-Hubbard models:
\begin{eqnarray*}
\mathcal{H}_\text{H}&=& J \sum_{i}\vec{S}_i\cdot\vec{S}_{i+1},\\
\mathcal{H}_\text{FH}&=& -t\sum_{i,\sigma} \left(c^{\dagger}_{i\sigma}c^{\vphantom\dagger}_{i+1\sigma}+h.c.\right)
+U\sum_{i} c^{\dagger}_{i\uparrow}c^{\vphantom\dagger}_{i\uparrow}c^{\dagger}_{i\downarrow}c^{\vphantom\dagger}_{i\downarrow}-\\
&&-\mu \sum_{i\sigma}  c^{\dagger}_{i\sigma}c^{\vphantom\dagger}_{i\sigma},
\end{eqnarray*}
where $\vec{S}$ are spin-$1/2$ operators and $ c^{\dagger}_\sigma,c_\sigma$ are spin-$1/2$ fermionic creation and annihilation operators respectively. For the Heisenberg model we use $J=1$, 
and for the Hubbard model we use $t=1$, $U=3$ and $\mu=U/10$. In both cases, the ground states are gapless in the thermodynamic limit with algebraically decaying correlation functions, although in practice we will simulate finite chains with open boundary conditions.

\subsection{Local Optimization versus Global Optimization }

We first compare local ``DMRG" style optimization versus global gradient-based optimization for the infidelity cost function ${\bar{\mathcal{F}}} = 1 - | \langle \Psi | \psi \rangle |$, where $| \Psi \rangle$ is the ground-state wavefunction of the model and $| \psi \rangle$ is the ansatz state. $| \Psi \rangle$ is obtained by the standard DMRG algorithm  using a dense MPS of sufficiently large bond dimension ($D \sim 400$) so that any error in $|\Psi\rangle$ is negligible. When $\bar{\mathcal{F}}=0$ then the circuit ansatz is identical to the ground-state wave function. 

The result of minimizing $\bar{\mathcal{F}}$ for  the different ansatz and optimization methods is shown in Fig.~\ref{fig:fidelOpt}. We show data from the quantum circuit MPS with a brick-wall local ansatz (qMPS-b) ($L=24$, $q=4$, $\tau=6$) and global brick-wall quantum circuit (QC-b) ($L=16$, $\tau=6$) as representative examples.
We find that in all cases, the local DMRG style optimization converges to the local minimum faster than the global gradient-based optimization using either the CG or L-BFGS-B algorithms. In addition, we find that in all cases, the L-BFGS-B algorithm converges more quickly than the CG algorithm. However, we also see that both the speed of convergence as well as the final converged result has some dependence on the initial guess. As observed in Fig.~\ref{fig:fidelOpt}, given a suitable initial guess, the  global gradient-based optimization eventually converges to a slightly lower minimum than that found by the local DMRG optimization. 

\subsection{Initial Guess}

The dependence of the optimization on the initial guess is well-known in quantum circuit optimization, where poor initial guesses can sometimes give rise to exponentially small gradients (the barren plateau problem~\cite{McClean:2018, Grant:2019,Tyler:2021, Cerezo:2021, Zhao:2021}). We can see a related problem in our circuits. To illustrate this, we show results from optimizing the energy cost function $E=\min \langle \psi | \mathcal{H} | \psi \rangle, \, |\psi \rangle \in \text{qTN}$ and we report the relative energy error $\delta E=E/E_\text{exact}-1$ versus iteration number in Fig.~\ref{fig:initrole} for several example circuits. We see in the top panel (Fig.~\ref{fig:initrole}(a)) that when starting from a random initial guess for the global QC-b ansatz,  we 
converge to the same relative error for two different circuit depths $\tau=8$ and $\tau=10$, despite the significantly larger number of variational parameters for $\tau=10$ versus $\tau=8$. 

To improve the initialization of larger circuits, we use optimized gates obtained from a shallow circuit to initialize gates at larger circuit depth.
The heuristic for this adaptive initialization method is summarized as follows:  (i) optimal gates are obtained from a random initial guess for a low depth circuit $\tau'$, (ii) the  initial guess for the ansatz with depth $\tau>\tau'$, is given by the optimized gates (from previous calculations) for $\tau \geq \tau-\tau'$ and the identity operator for $\tau< \tau - \tau'$, respectively (measuring depth from the register qubits) (see Fig.~\ref{fig:init} for an explicit example), (iii) small random perturbations to all gates in the ansatz are applied 
to avoid getting stuck in a local minimum, (iv) for a circuit ansatz with larger depth, we repeat steps (ii, iii). Empirically, it is found that a gradual increase of the circuit depth by $2-6$ layers works well, i.e. $\tau-\tau'=2-6$. The identity perturbation strength is also chosen to be of the same order as the local unitary gradient norm.

In Fig.~\ref{fig:initrole}(a), we show that optimizing the global QC-b ansatz with $\tau=10$ starting from optimal gates from $\tau=8$, indeed results in a lower relative error. In Fig.~\ref{fig:initrole}(b), we  plot the results of optimizing the QC-b ansatz with different $\tau$ (each initialized in the manner above, using lower depth circuits as shown in Fig.~\ref{fig:init}(c)), which shows that we can now achieve a systematic decrease in the relative error $\delta E$ as function of increasing $\tau$. Notice that in  all cases, it is necessary to first perform many iterations  to bring the ansatz out of the local minimum generated by the smaller $\tau$ guess, before one observes a significant drop in the relative error. 
 Similar results are shown for the qMPS-b ansatz with fixed $q=4$ and increasing $\tau$ for the local circuit (brick-wall ansatz) in Fig.~\ref{fig:initrole}(c);
 we similarly see that we can achieve a systematic decrease in the relative error when increasing $\tau$. 
Indeed, as we increase $\tau$ in the qMPS-b ansatz, we obtain results that approach the dense MPS (DMRG) result with bond dimension $D=2^q=16$. As this is the lower bound for the variational energy of any qMPS with $q=4$, our optimization heuristic using adaptive initialization thus fully realizes the variational power 
of the quantum circuit tensor network.


\begin{figure*}[ht!]
\centering
\includegraphics[width=1.0 \linewidth]{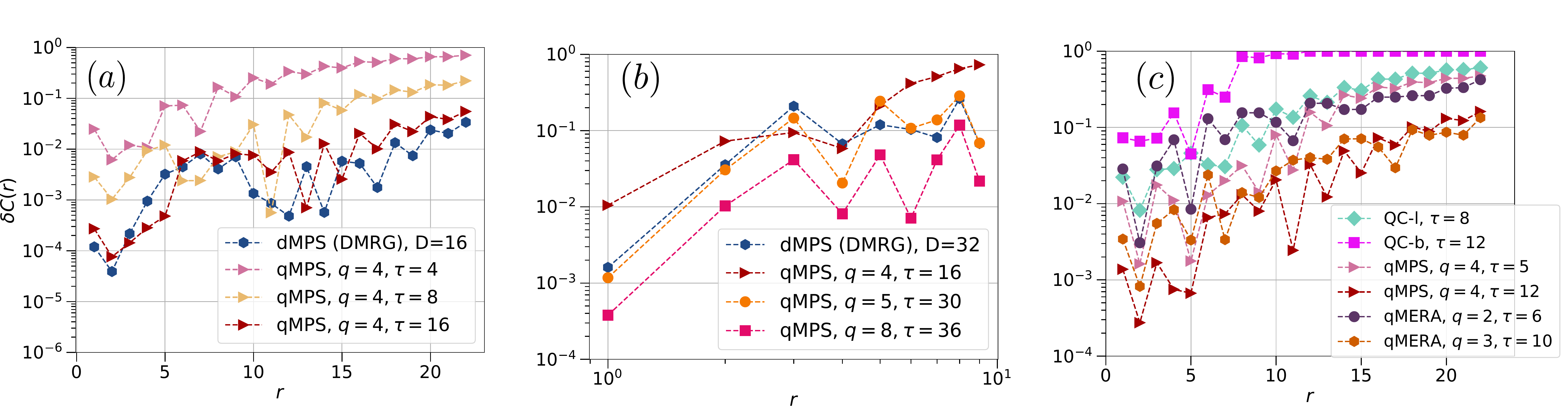}
\caption{ 
(Color online) Relative error in the spin-spin correlation function as a function of distance $r$.
(a), (c) are for the Heisenberg model, while (b) is for the Fermi-Hubbard model (both $L=32$). Both qMPS and qMERA use a local brick-work circuit (qMPS-b, qMERA-b). In (a), the qMPS with $q=4$, $\tau=16$ has fewer parameters than the dMPS with $D=16$, while in (b), the qMPS with $q=5$, $\tau=30$ has fewer parameters than the dMPS with $D=32$. In (c), the QC-l ansatz has a comparable numbers of parameters to the qMERA-b and qMPS-b with $q=2$, $\tau=6$ and $q=4$, $\tau=5$, respectively. qMPS-b ($q=4$, $\tau=12$) and qMERA-b ($q=3$, $\tau=10$) have comparable numbers of variational parameters. QC-b correlation functions are generally worse than those from the above ansatz.}
\label{fig:corr}
\end{figure*}

\section{Variational power of quantum circuit tensor networks, dense tensor networks, and global quantum circuits}

\subsection{Energies}

Using the above optimization strategies, we can now systematically characterize the variational power of the different ans\"atze discussed in this work for ground-state representation. As the measure of expressiveness, we use the relative energy error $\delta E$ as a function of the number of variational parameters. (Related measures have been recently used to compare different variational wavefunctions~\cite{stair2020exploring}). The various ansatz are optimized using the global gradient scheme with the L-BFGS-B algorithm, while the dense MPS results are obtained by the DMRG algorithm. The largest parametrized circuit ans\"atze correspond to the following: (i) for qMPS-b and qMERA-b, the number of bond qubits and local circuit depths are $q=8, \tau=32$ and $q=4, \tau=8$ respectively, and (ii) for the global circuit ansatz QC-b and QC-l, the largest circuit depths are $\tau=14$ and $\tau=9$ respectively. 
Despite the large number of circuit parameters, we find that the optimization heuristics work to high accuracy, if sufficient iterations are used. For example, the smallest relative energy errors we find using the qMPS ansatz are $10^{-6}$ and $10^{-4}$ for the Heisenberg and Fermi-Hubbard models with $L=32$, respectively, using   $\sim 10^{6-7}$ iterations.

\begin{table}[t]
\begin{tabular}{|c|c|c|c|c|}
\hline  Ansatz & $\text{Heisenberg}$, $(a, b)$ &$\text{Hubbard}$, $(a, b)$  \\ \hline
qMPS-b &$(20, 4.0)$&$(9,  1.9)$  \\
qMPS-l &$(14, 3.1)$&$(10,  1.9)$ \\
QC-b &$(4 , 1.4)$ &$(4.4, 1.0)$  \\ 
QC-l &$(8, 2.2)$&$(0.4,   0.5)$  \\ 
qMERA-b &$(15, 3.1)$&$(6.0, 1.4)$  \\ 
dMPS (DMRG) &$(15, 2.9)$&$(8.0, 1.5)$  \\ 
dMERA &$(3.5, 1.2)$&$(0.8, 0.6)$  \\ \hline
\end{tabular}
\caption{Scaling coefficients $(a, b)$ in the form $\delta E \sim an^{-b}$  for the various ansatz in the Heisenberg and Fermi-Hubbard models. The asymptotic behaviour of the relative error $\delta E$ at large $n$ is controlled by $b$.}
\label{tab:fresult}
\end{table}

We benchmark the performance of the qMPS, qMERA, and global QC ansatz versus the dense MPS (DMRG) and dense MERA for the 1D Heisenberg and Fermi-Hubbard models with $L=32$ (2D results are discussed in a later section). The key findings are as follows: (i) Comparing dense MPS with qMPS, we find that for an equivalent number of variational parameters, qMPS achieves lower energies than the dense MPS (see Figs.~\ref{fig:deltaE}(a), (d)) in both the Heisenberg and Fermi-Hubbard models. (ii) Similarly, comparing dense MERA with qMERA, we  find that for an equivalent number of variational parameters, qMERA achieves lower energies than dense MERA. Taken together with the previous statement, this implies that the \emph{appropriate quantum circuit tensor networks are more compact and expressive than their traditional dense counterparts} for these problems. In particular, the worst case possibility, that one requires an exponential number of gates to accurately parametrize the local block unitary, does not apply to these physical ground states. (iii) Such expressiveness is not shared by the global brick-wall and ladder circuits (Figs.~\ref{fig:deltaE}(b), (e)) which are consistently less expressive than the qMPS.
This is consistent with the theoretical analysis in Section~\ref{sec:analysis}, which identifies QC-b and QC-l as constrained versions of the qMPS with minimal parametrizations of the block unitaries. (iv) qMPS is somewhat more expressive than qMERA (Figs.~\ref{fig:deltaE}(c), (f)). This suggests that the formal ability to capture logarithmic corrections (which exist in the thermodynamic limit of the 1D Heisenberg model) is either unimportant for the energy or at the system size considered.

Empirically, we can also summarize the data by fitting the relative error 
to the inverse polynomial $\delta E(n) \sim an^{-b}$, where $n$ is the number of variational parameters. As shown in Fig.~\ref{fig:deltaE} in the log-log plot, this form fits all the ans\"atze reasonably well, with some small systematic deviations, for example in the case of DMRG at larger $n$. 
A linear fit  to the log-log data yields 
an estimate of $a$ and $b$, as shown in Table~\ref{tab:fresult}, where $b$ gives the asymptotic scaling for large $n$. 
These results further support the findings above: in the large $n$-limit, the qMPS ansatz is the most expressive ansatz and outperforms the dense MPS, while qMERA outperforms the dense MERA, with both also outperforming the global brick-wall and ladder ansatz. 
We  see that the brick-wall local circuit structure for qMPS yields
a better overall performance than local ladder/MERA structures (see also Appendix~\ref{sec:appendixb}). The qMERA ansatz performs similarly to the dense MPS algorithm in both models with $b_\text{qMERA} \approx b_\text{dMPS}$. The ratio $b_{\text{qMPS}}/b_\text{qMERA}$ and $b_{\text{qMPS}}/b_\text{dMPS}$ is $\sim 1.3$ in both models. Earlier studies of scale-invariant MERA and infinite MPS have found a similar ratio, $b_{\text{MPS}}/b_\text{MERA} \sim 1.2$ \cite{Avella:2013}. 
Finally, depending on the model, either QC-b or QC-l performs better:  $b_{\text{QC-b}} < b_{\text{QC-l}}$ ($b_{\text{QC-b}} > b_{\text{QC-l}}$) for the Heisenberg (Fermi-Hubbard) models, respectively.

\subsection{Correlation functions}

We next study how faithfully the different ans\"atze capture correlation functions of the Heisenberg and Fermi-Hubbard models. We use the spin-spin correlation function,  defined as  $\langle \vec{S}_{0}\cdot\vec{S}_{r}\rangle-\langle \vec{S}_0^2\rangle$ (Heisenberg and Fermi-Hubbard) 
as a representative example.  

In Fig.~\ref{fig:corr}, we show errors in the correlation function $\delta C(r)$ (relative to numerically exact data) for the various ans\"atze.  In Fig.~\ref{fig:corr}(a) (Heisenberg model), we see that qMPS-b with $q=4, \tau=16$ produces the same algebraically decaying correlation function as a dense MPS with $D=16$  over the full distance range of  $r<22$, despite having fewer variational parameters. In Fig.~\ref{fig:corr}(b) (Hubbard model), we similarly find that qMPS-b with $q=5, \tau=30$ produces a similar quality correlation function to the dense MPS with $D=32$, but again with fewer variational parameters. These results are consistent with the greater expressiveness of the quantum circuit tensor network relative to its dense counterpart. 

In both models, we see that increasing either $\tau$ or $q$ leads to an improvement of the correlation function. However, neither is a dominant factor for convergence. For example, in the Heisenberg model, we find that using a qMPS-b with $q=5, \tau=14$ provides a lower relative error compared to $q=4, \tau=18$, despite having fewer variational parameters; but a qMPS-b with $q=8, \tau=4$  performs similarly to $q=4, \tau=5$, despite having a large number of variational parameters. Thus a balanced choice of $q, \tau$ is needed to obtain the best result.

In Fig.~\ref{fig:corr}(c), we show the qMERA-b, QC-l and QC-b correlation functions alongside the qMPS-b correlation functions. One expects that qMPS-b will accurately reproduce short-range correlations (up to the MPS correlation length)
while  qMERA should perform better at long distances. 
Quantitatively, we find that qMPS-b ($q=4, \tau=12$ and $\delta E=6\times 10^{-5}$) indeed provides a lower relative error at short distances, while qMERA-b (with $q=3, \tau=10$ and $\delta E=3\times10^{-4}$) with a similar number of variational parameters is more accurate at longer distances ($r>15$). In addition, qMPS with $q=4, \tau=5$ outperforms QC-l with a similar number of parameters, especially at short distances.  At long range,  qMERA-b with $q=2, \tau=6$ is clearly better than QC-l, while at short distances it is comparable. Overall, qMPS and qMERA thus appear to provide more faithful representations of the correlation functions than the general quantum circuit ansatz, again reinforcing the need to balance the spreading of entanglement and the accuracy of the local block unitary representation. In addition, the improved entanglement spreading structure of the qMERA is detected in the correlation functions, even though it is not represented in the energy metric of the previous section.

%

\begin{figure}
\begin{center}
\includegraphics[width=1.0 \linewidth]{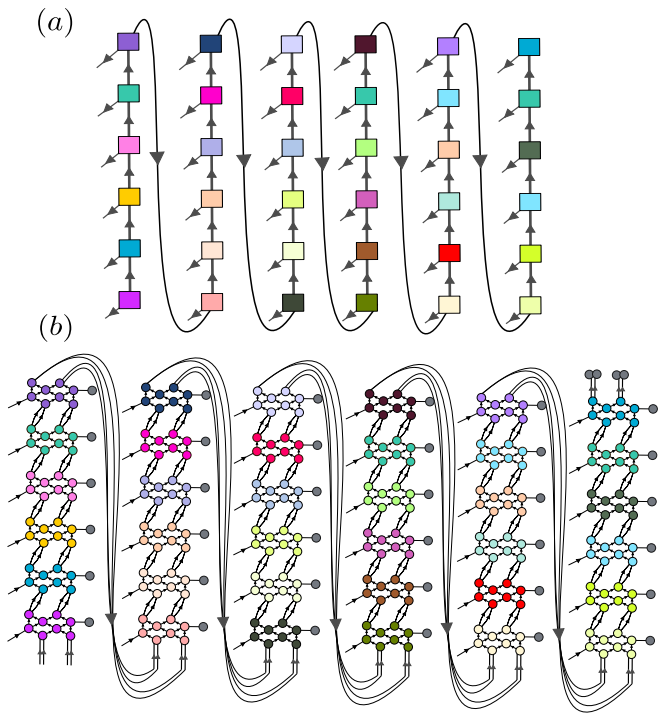} 
  \caption{(Color online) Schematic representation of (a) a snake dMPS and (b) its quantum circuit (qMPS) used in the two-dimensional simulation of Heisenberg model with $L=6 \times 6$ sites. The qMPS-b shown has four bond qubits $q=4$ with depth $\tau=4$.}
  \label{fig:2dqmps}
\end{center}
\end{figure}

\begin{figure}
\begin{center}
\includegraphics[width=.90 \linewidth]{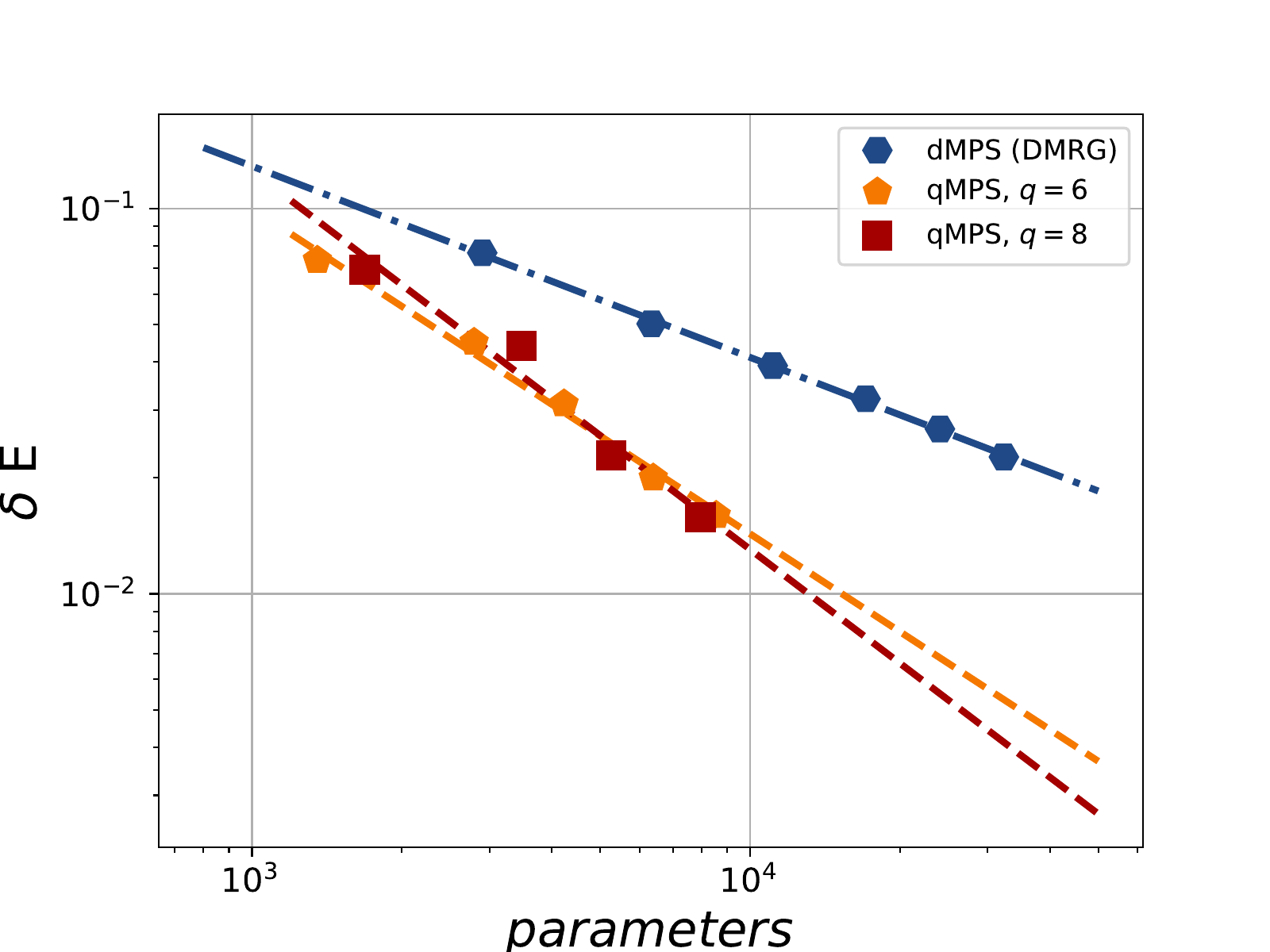} 
  \caption{(Color online) Comparison between qMPS and dMPS for the two-dimensional Heisenberg model on a $6 \times 6$ lattice.}
  \label{fig:2dqmps}
\end{center}
\end{figure}

\begin{table}[t]
\begin{tabular}{|c|c|c|c|c|}
\hline  Ansatz & $\text{2d Heisenberg}$, $(a, b)$  \\ \hline
qMPS-b &$(4.7, 1.1)$  \\
dMPS (DMRG) &$(1.4, 0.48)$  \\  \hline
\end{tabular}
\caption{Scaling coefficients $(a, b)$ for qMPS and dMPS in the two-dimensional Heisenberg model on a $6 \times 6$ lattice.}
\label{tab:2Dresult}
\end{table}

\subsection{Two dimensional systems}

We have also carried out a limited set of studies on a two-dimensional Heisenberg model using the qMPS ansatz arrange in a traditional snake through the two-dimensional lattice as depicted in Fig.~\ref{fig:2dqmps}.
In two-dimensional systems, it is well known that a traditional dense MPS (arranged as a snake) requires an exponentially large bond dimension in the system width to represent area law ground states. Since states with exponentially large bond dimension can be reached by qMPS with a polynomial circuit depth, one might expect a larger expressivity difference between qMPS and dMPS in 2D. The relative energy error $\delta E$ is plotted as a function of the number of parameters in Fig.~\ref{fig:2dqmps}, and the corresponding fit of the relative error $\delta E(n)\sim an^{-b}$ is shown in Table~\ref{tab:2Dresult}. From both of these, we indeed find that the expressivity gap between qMPS and dMPS is greater in 2D, with $b_\text{qMPS}/b_\text{dMPS}=1.4$ in 1D, but $b_\text{qMPS}/b_\text{dMPS}=2.3$ in 2D. Of course, in two and more dimensions, the possible choices of circuit architecture are richer, with many types of isometric tensor networks to explore beyond the qMPS~\cite{Haghshenas:2019, Zaletel:2020,slattery2021quantum,maccormack2021simulating}.

\section{Discussion of quantum advantage}
Our numerical results clearly show that for ground states of some commonly considered physical models, there is a quantum advantage in the expressivity of quantum circuit tensor networks versus the traditionally employed classical dense tensor networks. The difference in scaling of the
achieved accuracy  as a function of the number of parameters suggests that the advantage in expressiveness will persist into a regime where the circuits can no longer be contracted efficiently classically. Assuming standard tensor networks are the best classical representation for these problems, this means that in the high accuracy regime, quantum circuit tensor networks may have the (theoretical) potential to achieve quantum advantage also in terms of computational cost.

To explicitly translate the advantage in representation to one of computation, we must consider the cost to compute with the quantum circuit representation (on a quantum device) versus the dense representation on a classical device. We consider the case of MPS as an example. The cost to contract a classical dense MPS is $O(D^3)$, or $O(n^{3/2})$, where $n$ here denotes the number of local parameters in the block unitary. 
(We ignore scaling with $L$ here and below). 
For qMPS, the cost to run the circuit to prepare the state (assuming gates are executed sequentially) is $O(n)$. We then imagine computing the energy by sampling terms in the Hamiltonian; for a relative precision $\delta E$, we require $O(1/(\delta E)^2)$ measurements per term.
Combining these factors together with the empirical scaling of $\delta E$ with $n$, one finds that the cost $T$ to compute the energy 
to an accuracy of $\delta E$ is $T\sim \delta E^{0.52}$ (classical dMPS) and $T\sim \delta E^{2.25}$ (qMPS-b) in the 1D Heisenberg model, and $T\sim \delta E^{3.1}$ (classical dMPS) and $T\sim \delta E^{2.9}$ (qMPS-b) in the 2D Heisenberg model. 
A similar analysis for MERA finds in the 1D Hubbard model, $T\sim \delta E^{3.75}$ (classical dMERA) and $T\sim \delta E^{2.7}$ (qMERA). 
These small polynomial advantages (where they appear) are perhaps reflective of the challenges of variational quantum algorithms, and whether they are realizable, or persist with improved classical techniques remains to be seen. However, it should be noted that the asymptotic inefficiency of the quantum algorithm stems from the steep cost associated with sampling expectation values. Techniques that trade coherence for reduced sampling, for example reducing the number of repetitions to as few as $O(\log(1/(\delta E)))$, with a measurement circuit depth proportional to $O(1/(\delta E))$, therefore greatly affect this analysis of computational advantage~\cite{wang2019accelerated}. 

Of course to seize this potential advantage, one would also need to optimize circuit parameters in this classically-intractable regime. In the case of gradient optimization, classical algorithms obtain the gradient at the same cost as the energy through backpropagation, but for quantum algorithms using finite differences (for example using the parameter shift rule)~\cite{mitarai2018quantum,Banchi2021measuringanalytic}, the energy evaluation must be repeated $O(n)$ times. In this case, the above polynomial advantages will disappear unless coherent expectation value techniques are used, which may be further combined with coherent techniques for gradient evaluation~\cite{gilyen2019optimizing}.
Also, we have assumed that the number of optimization iterations needed to find the ground-state scales with $\delta E$ in a comparable way in the quantum and classical computations. 
While the optimization heuristics discussed in this work successfully find accurate ground-states with a tractable number of optimization iterations, it remains to be seen whether this scaling persists in very large circuits.




\section{Conclusions}
\label{Sec:CONCLUSION}

In this work we studied the variational power of quantum circuit tensor networks, and in particular, quantum circuit matrix product states and the quantum circuit multi-scale entanglement renormalization ansatz, for representing the ground-states of quantum many-particle problems. As we argued, this is a problem where standard tensor networks excel, and is thus a high bar for quantum circuit tensor networks to meet. 
We found that quantum circuit tensor networks outperform other common global quantum circuit ansatz in variational power, requiring far fewer parametrized gates for a given accuracy. In fact, they appear to be
asymptotically even more expressive than the standard tensor networks, in terms of the number of parameters to converge to a comparable accuracy in the variational energy and correlation functions. Our initial results in 2D suggest that this expressive advantage increases in higher dimensions.

Although all simulations here were carried out classically, the difference in expressiveness of the classical and quantum circuit tensor networks raises the possibility of polynomial quantum advantage in the computation of certain problems. The practical realization of such advantage critically depends both on the performance of optimization heuristics (such as the one proposed here) as well as the cost of estimating expectation values on quantum devices. However, the variational power of the quantum circuit tensor networks provides new motivation to improve the optimization strategies for this class of circuits. It also provides impetus to study related types of ansatz in the context of classical simulations, where they may provide the chance to improve on long-standing tensor network paradigms.

\acknowledgments
Primary funding for this work (RH, GKC) was provided by the US Department of Energy, Office of Science, via DE-SC0019374. GKC acknowledges additional support from the Simons Foundation via
the Many-Electron Collaboration and via the Simons Investigator program. ACP was supported by the US NSF Convergence Accelerator Track C award 2040549. The numerical codes were implemented in the  \texttt{quimb} library~\cite{gray2018quimb} which is freely available~\cite{quimbgithub, qMPSgithub}. Support for JG and the development of the \texttt{quimb} library was provided by a gift from Amazon Web Services, Inc.

\bibliography{Ref}

\begin{thebibliography}{53}
\bibitem [{mer()}]{mera:note}%
  \BibitemOpen
  \href@noop {} {}\bibinfo {note} {Empirically, we find that qMERA with local
  brick-wall and ladder circuits variationally behave very similarly, at least
  within the range of accessible qubits $q$ and circuit depth $\tau$ presented
  in this work.}\BibitemShut {Stop}%
\bibitem [{cou()}]{counting:note}%
  \BibitemOpen
  \href@noop {} {}\bibinfo {note} {A further reduction in the number of
  parameters is possible by eliminating one parameter along the connected
  bonds, absorbing a rotational gate into adjacent gates, and also $6$
  parameters for two gates acting on top of each other. In our reported data,
  we have considered these reductions, which mainly change the scaling
  prefactor $a$, not the important exponent $b$.}\BibitemShut {Stop}%
\end{thebibliography}

\begin{figure}
\begin{center}
\includegraphics[width=.90 \linewidth]{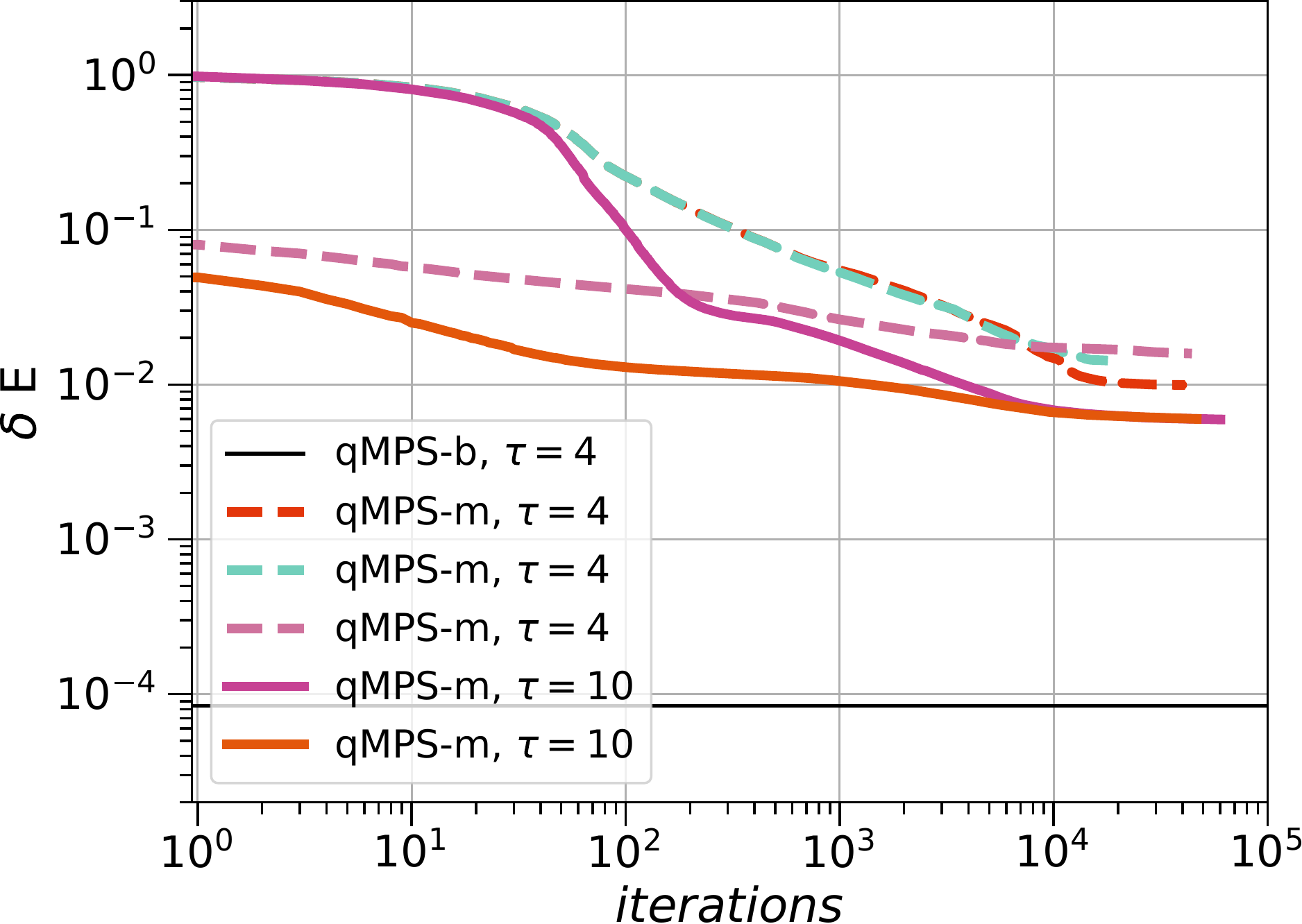} 
  \caption{(Color online) Comparison between qMPS with MERA and brick-wall local circuits. The solid and dashed lines represent two different initial starting states.  }
  \label{fig:qmpsMERA}
\end{center}
\end{figure}

\appendix

\section{qMPS with MERA local circuits}
\label{sec:appendixb}
We benchmark the accuracy of the qMPS-m ansatz by studying the relative error $\delta E$ for the Heisenberg model $\mathcal{H}_{\text{H}}$. In general, a systematic study of the ansatz is difficult as there are many controlling parameters: $q, q_m, \tau$ where $q_m$ is the number of bond qubits for the local MERA; and in addition, there is also the choice of the internal structure of the local MERA. To simplify things, we fix the number of bond qubits to $q=8$ and use a brick-wall structure for the MERA with $q_m=3$. Empirically, we find that it is difficult to obtain converged results for qMPS-m as it easily gets stuck in local minima. Thus, for the reported data, two different initial states were chosen, one random and one obtained from a smaller optimized circuit. In Fig.~\ref{fig:qmpsMERA}(a), we compare qMPS-m with qMPS-b for different $\tau$. We find that the qMPS-b with $\tau=4$ easily outperforms qMPS-m with $\tau=4$ (which has a larger number of variational parameters). We see that increasing $\tau$ in qMPS-m from $4$ to $10$ only slightly improves the relative error, i.e. from $9\times 10^{-3}$ to $6\times 10^{-3}$. Overall, the complexity of this circuit structure may require additional improvements in optimization strategy in order to realize its variational power.

\end{document}